\documentclass[journal,final,letterpaper]{IEEEtran}

\IEEEoverridecommandlockouts

\usepackage{cite}
\usepackage{amsfonts}
\usepackage[cmex10,nosumlimits]{amsmath}
\usepackage{graphicx}
\usepackage[usenames,dvipsnames]{pstricks}
\usepackage{algorithm}
\usepackage[noend]{algpseudocode}
\usepackage[font=normalsize]{subfig}

\usepackage{flushend}

\newtheorem{definition}{Definition}
\newtheorem{assumption}{Assumption}

\newtheorem{theorem}{Theorem}
\newtheorem{lemma}{Lemma}
\newtheorem{proposition}{Proposition}
\newtheorem{corollary}{Corollary}

\DeclareMathOperator*{\maximize}{maximize}

\newcommand{\pp}[1]{{\left( #1 \right)}}
\newcommand{\br}[1]{{\left\{ #1 \right\}}}
\newcommand{\ebr}[1]{{\left[ #1 \right]}}

\newcommand{\abs}[1]{{ | #1 | }}
\newcommand{\sabs}[1]{{ | #1 |^2 }}

\newcommand{\mat}[1]{\boldsymbol{#1}}

\def\setA{{\mathcal{A}}} % beamforming vector space
\def\setB{{\mathcal{B}}}
\def\setD{{\mathcal{D}}}

\def\setL{{\mathcal{L}}}

\def\setK{{\mathcal{K}}}
\def\setX{{\mathcal{X}}}
\def\setS{{\mathcal{S}}}
\def\setZ{{\mathcal{Z}}}

\def\setR{{\mathcal{R}}}

\def\su{\text{\scriptsize{su}}} % beamforming vector space
\def\susum{\text{\scriptsize{su-sum}}} % beamforming vector space
\def\pu{\text{\scriptsize{pu}}} % beamforming vector space

\begin{document}
\title{Distributed Channel Assignment in Cognitive Radio Networks: Stable Matching and Walrasian Equilibrium}

\author{Rami~Mochaourab,
        Bernd~Holfeld,
        and~Thomas~Wirth% <-this % stops a space
        %
       %\thanks{\copyright \copyright~2014 IEEE. Personal use of this material is permitted. Permission from IEEE must be obtained for all other uses, in any current or future media, including reprinting/republishing this material for advertising or promotional purposes, creating new collective works, for resale or redistribution to servers or lists, or reuse of any copyrighted component of this work in other works.}%
        \thanks{This work has been submitted to the IEEE for possible publication. Copyright may be transferred without notice, after which this version may no longer be accessible.}
        \thanks{Rami Mochaourab was with Fraunhofer Heinrich Hertz Institute, Berlin, Germany. Now he is with ACCESS Linnaeus Centre, Signal Processing Department, School of Electrical Engineering, KTH Royal Institute of Technology, 100 44 Stockholm, Sweden. Phone: +4687908434. Fax: +4687907260. E-mail: rami.mochaourab@ieee.org. Bernd Holfeld and Thomas Wirth are with Fraunhofer Heinrich Hertz Institute, Berlin, Germany. E-mail: bernd.holfeld@hhi.fraunhofer.de, thomas.wirth@hhi.fraunhofer.de.}% <-this %
        %
        %\thanks{This work was funded by the Federal Ministry of Education and Research (BMBF) of Germany in the framework of the Cognitive Mobile Radio (CoMoRa) project under support grant 16BU1200.}
        }%

\maketitle

\begin{abstract}
We consider a set of secondary transmitter-receiver pairs in a cognitive radio setting. Based on channel sensing and access performances, we consider the problem of assigning channels orthogonally to secondary users through distributed coordination and cooperation algorithms. Two economic models are applied for this purpose: matching markets and competitive markets. In the matching market model, secondary users and channels build two agent sets. We implement a stable matching algorithm in which each secondary user, based on his achievable rate, proposes to the coordinator to be matched with desirable channels. The coordinator accepts or rejects the proposals based on the channel preferences which depend on interference from the secondary user. The coordination algorithm is of low complexity and can adapt to network dynamics. In the competitive market model, channels are associated with prices and secondary users are endowed with monetary budget. Each secondary user, based on his utility function and current channel prices, demands a set of channels. A Walrasian equilibrium maximizes the sum utility and equates the channel demand to their supply. We prove the existence of Walrasian equilibrium and propose a cooperative mechanism to reach it. The performance and complexity of the proposed solutions are illustrated by numerical simulations.
\end{abstract}%The mechanism is a modification of an English auction and requires direct communication between the secondary users. 

\IEEEpeerreviewmaketitle
\section{Introduction}%
In cognitive radio settings, secondary users (SUs) are capable of adapting their transmissions intelligently \cite{Haykin2005}. Through the detection of spectrum holes, the SUs can use the unoccupied channels licensed to the primary users for communication. This mechanism is called opportunistic spectrum access \cite{Zhao2007} and corresponds to the interweave paradigm described in \cite{Goldsmith2009}.

Generally, there exists a tradeoff between the optimization of the secondary systems' performance and the primary systems' performance \cite{Liang2008}. Our objective is to find an assignment of the primary channels to the SUs taking into account both secondary and primary user performances. For a survey on channel assignment mechanisms in cognitive radio networks please refer to \cite{Tragos2013}. Since a cognitive radio network is a distributed and less regulated system, we are interested in channel assignment mechanisms which are implemented in a distributed way. We study such mechanisms using matching markets and competitive markets with indivisible goods.

Although the applications and solutions of the two market models are conceptually different, there exist similarities between the two market models \cite{Jackson13}. First, both solutions of the market models lead to an assignment, which is in our case, an orthogonal assignment of the channels to the SUs. Moreover, both models assume autonomous and rational agents who are able to decide locally between different alternatives. These properties are favorable for distributed operation of the SUs. Nevertheless, both models, rely on communication based on binary decisions reflecting a proposal in stable matching or a demand in competitive markets. Hence, the application of the two models has practical implementation in cognitive radio networks in which coordination can be achieved with low communication overhead. In addition to their distributed and low communication overhead properties, optimality of the solutions of both frameworks within specified performance regions make the application of these models attractive for resource allocation in communication networks. We relate to some of these works, after discussing the differences between the two frameworks.

The differences between the two models are as follows: Competitive markets use prices as means to coordinate the demands (decisions) of the consumers to buy goods and are updated by an auction mechanism to reach a solution. In stable matching, on the other hand, no prices are involved but the two sets of agents, i.e. SUs and channels, exchange proposals based on preference relations of each agent within the two sets. Through sequences of acceptances and rejections, a stable matching is reached. In the competitive market model, only the consumers' preferences (utility functions) are needed. In Section \ref{sec:problem_def}, we further discuss the differences of the two solutions for our cognitive radio scenario.%
\subsection{Application of Matching and Competitive Market Models}%
%%%%%%%%%%%%%%%%%%%%%%%%%%
%
In two-sided matching markets \cite{Roth1990}, two sets of agents are to be matched, corresponding to the SUs and the primary channels. Each agent in one set has preferences over the agents in the other set. A matching of the agents in the two sets is stable when no pairs of agents prefer each other compared to their current matching.

Matching market models for resource allocation in wireless networks have been recently applied in several works. In \cite{Jorswieck2011}, the framework of two-sided stable matching is applied for resource allocation in wireless networks and its merits revealed regarding distributed implementation and efficiency. Stable matching for channel assignment in cognitive radio settings has been applied in \cite{Yaffe2010,Leshem2012,Huang2013,Jorswieck2013}. In \cite{Yaffe2010} and \cite{Leshem2012}, one-to-one stable matching is considered where the utility of the secondary and primary users are chosen to be identical due to the fact that the SUs cannot obtain the performance measures of the primary users. In this case, the stable matching of SUs to the primary channels is proven to be unique. In addition, in \cite{Leshem2012} stable matching is successfully implemented through opportunistic CSMA techniques. Reference \cite{Huang2013} applies the model in \cite{Leshem2012} to interweave cognitive radio settings with identical utility for secondary and primary users. While in \cite{Leshem2012,Huang2013}, the utility of both types of agents are the same, in \cite{Jorswieck2013} the utility of the primary users depend on the interference leakage from SUs and the utility of the SUs are their achievable rates in the primary channels. In this context, many-to-one stable matching is applied.

Stable matching for channel assignment in a single radio cell is applied in \cite{El-Hajj2012} where two-sided matching takes into account the utilities of the users in the uplink and the downlink transmissions. In \cite{Holfeld2013}, the stable matching framework is applied for cross-layer scheduling in the downlink of a single cell where the utility of a user is his sum rate and the utility of the resources includes the user queue state of the buffer. In the context of physical layer security, stable matching of transmitter-receiver pairs to friendly jammer is proposed in \cite{Bayat2013}. In \cite{Saad2014}, uplink user association in small cell networks is considered using many-to-one stable matching as well as coalitional games. The user utilities are based on quality of service (QoS) and coverage aspects.

%
%
%\subsection{Application of Competitive Market Models and Auctions}%
%%%%%%%%%%%%%%%%%%%%%%%%%%%%%%%%%%
%
In competitive markets \cite{Jehle2003}, also referred to as one-sided matching markets \cite{Kelso1982}, there exists a set of agents which want to buy quantities of goods. The prices of the goods regulate the quantities bought by the consumers and are adapted depending on the demand and supply of the goods. The Walrasian equilibrium is a state in which the demand equals the goods' supply. In order to reach a Walrasian equilibrium, a price adjustment process is required. This process is related to auction mechanisms and its advantage is the distributed implementation aspect and the limited amount of information exchange required between the users and the coordinator. 

Competitive market models have found a few applications for resource allocation problems in communication networks. Please refer to \cite{Berry2010,Jorswieck2013a} for a discussion on these applications. Also, for a recent survey on auction mechanisms for resource allocation in wireless networks, see \cite{Zhang2013}. In cognitive radio settings, auctions have been applied for distributed channel assignment in \cite{Han2011,Naparstek2014a,Naparstek2014}. In \cite{Han2011}, repeated auctions in the uplink of a secondary cell are proposed for the allocation of the primary channel resources to the SUs. Distributed auctions are studied in \cite{Naparstek2014a} for energy efficient channel assignment in cognitive radios. Moreover, in \cite{Naparstek2014}, a distributed auction mechanism is proposed to find optimal one-to-one channel assignment to the SUs where CSMA mechanisms are utilized to implement the solution.% The framework is however restricted to assigning each secondary user a single channel.%
\subsection{Contributions and Outline}%
%%%%%%%%%%%%%%%%%%%
%
In this work, we consider a set of transmitter-receiver pairs as SUs and each user seeks the assignment of a set of primary channels. Our objective, formulated in Section~\ref{sec:problem_def}, is to optimize both the secondary and primary users' performance through coordinated and cooperative distributed channel assignment. We assume that each primary channel can be assigned to one SU while an SU can be assigned to multiple channels. However, an SU is restricted to use a maximum number of channels called \emph{quota} which improves the fairness in the channel assignment. 

We propose a coordinated channel assignment (Section~\ref{sec:stable_matching} ) which exploits many-to-one stable matching. Here, we assume that a coordinator exists which can communicate with the SUs. We characterise in worst case the number of bits each SU has to exchange with the coordinator in order to reach a stable matching. In addition, we provide conditions under which the stable matching is unique and primary user optimal. Our model differs from the models used in \cite{Leshem2012,Huang2013}, by the following two aspects: multiple channels are assigned per SU, and the utility functions of the primary channels are different from the utility functions of the SU. One main difference to \cite{Jorswieck2013} is our application of stable matching in interweave cognitive radio.

For cooperative channel assignment, we study a competitive market model with indivisible goods \cite{Gul1999} in Section~\ref{sec:walras_equilibrium}. The utility function of an SU is the weighted sum of his achievable rate and the utility of the primary users whose channels he is assigned to. We prove the existence of a Walrasian equilibrium which maximizes the weighted sum-performance of the secondary and primary systems. To reach the Walrasian equilibrium through a cooperative mechanism, we exploit an English auction algorithm from \cite{Gul2000}. The cooperative mechanism requires the exchange of $L$ bit information between the SUs. In comparison to auction algorithm studied in \cite{Naparstek2014}, our mechanism is able to assign multiple channels to each user.%

Numerical simulations are provided in Section~\ref{sec:simulations} before we draw the conclusions in Section~\ref{sec:conclusions}.%
\subsubsection*{Notations}%
Vectors are written in boldface letters. Sets are written in calligraphic font. $\abs{\setS}$ is the cardinality of the set $\setS$. $\abs{c}$ is the absolute value of $c \in \mathbb{C}$. The Q-function is given as $Q(x) = \frac{1}{\sqrt{2\pi}} \int_x^\infty \exp(-{u^2}/{2}) du$. The inverse of the Q-function is $Q^{-1}(x)$. $x \sim \mathcal{CN}(0,a)$ is a circularly-symmetric Gaussian complex random variable with zero mean and variance $a$. $\text{Pr}(x)$ is the probability of an event $x$. $\mathbb{R}_+$ is the set of nonnegative real numbers.%
%
% Interference temp constraint \cite{Zhang2010}
%\cite{Ye2007,Lin2009,Xie2010}
%\cite{Wang2008}
\section{System Model}\label{sec:sys_model}
\begin{figure}
  \centering
  % \usepackage[usenames,dvipsnames]{pstricks}
% \usepackage{epsfig}
% \usepackage{pst-grad} % For gradients
% \usepackage{pst-plot} % For axes
% User Packages:
% 
% \usepackage{amsfonts}
\begin{center}
\psscalebox{1.0 1.0} % Change this value to rescale the drawing.
{
\begin{pspicture}(0,-1.9954687)(8.52,1.9954687)
\pscircle[linecolor=black, linewidth=0.04, dimen=outer](1.95,1.3854687){0.15}
\rput[br](3.5,1.7354687){\footnotesize{primary transmitter $l\in \mathcal{L}$}}
\pscircle[linecolor=black, linewidth=0.04, dimen=outer](6.85,1.3854687){0.15}
\rput[bl](5.5,1.7354687){\footnotesize{primary receiver $l \in \mathcal{L}$}}
\pscircle[linecolor=black, linewidth=0.04, dimen=outer, doubleline=true, doublesep=0.02](1.95,-1.3848437){0.15}
\rput[br](3.7,-1.9954687){\footnotesize{secondary transmitter $k \in \mathcal{K}$}}
\pscircle[linecolor=black, linewidth=0.04, dimen=outer, doubleline=true, doublesep=0.02](6.85,-1.3848437){0.15}
\rput[bl](5.3,-1.9954687){\footnotesize{secondary receiver $k \in \mathcal{K}$}}
\psline[linecolor=black, linewidth=0.04, arrowsize=0.04cm 2.0,arrowlength=2.0,arrowinset=0.0]{->}(2.4,1.3354688)(6.5,1.3354688)
\psline[linecolor=black, linewidth=0.04, doubleline=true, doublesep=0.02, arrowsize=0.039999999999999994cm 2.0,arrowlength=2.0,arrowinset=0.0]{->}(2.4,-1.3645313)(6.5,-1.3645313)
\psline[linecolor=black, linewidth=0.04, doubleline=true, doublesep=0.02, arrowsize=0.039999999999999994cm 2.0,arrowlength=2.0,arrowinset=0.0]{->}(2.4,-1.1645312)(6.5,1.1354687)
\psline[linecolor=black, linewidth=0.04, arrowsize=0.04cm 2.0,arrowlength=2.0,arrowinset=0.0]{->}(2.4,1.1354687)(6.5,-1.2645313)
\psline[linecolor=black, linewidth=0.04, arrowsize=0.04cm 2.0,arrowlength=2.0,arrowinset=0.0]{->}(1.9,1.0354687)(1.9,-1.0645312)
\rput[bl](4.1,1.1354687){\footnotesize{\psframebox*[framesep=0, boxsep=false,fillcolor=white] {$g^{[l]}$}}}
\rput[bl](4.1,-1.5645312){\footnotesize{\psframebox*[framesep=0, boxsep=false,fillcolor=white] {$h_{k}^{[l]}$}}}
\rput[bl](3.3,0.33546874){\footnotesize{\psframebox*[framesep=0, boxsep=false,fillcolor=white] {$\tilde{g}^{[l]}$}}}
\rput[bl](3.3,-0.76453125){\footnotesize{\psframebox*[framesep=0, boxsep=false,fillcolor=white] {$\tilde{h}_{k}^{[l]}$}}}
\rput[bl](1.7,-0.16453125){\footnotesize{\psframebox*[framesep=0, boxsep=false,fillcolor=white] {$z_{k}^{[l]}$}}}
\end{pspicture}
}
\end{center}
  \caption{Illustration of the system model.}\label{fig:sysmodel}
\end{figure}
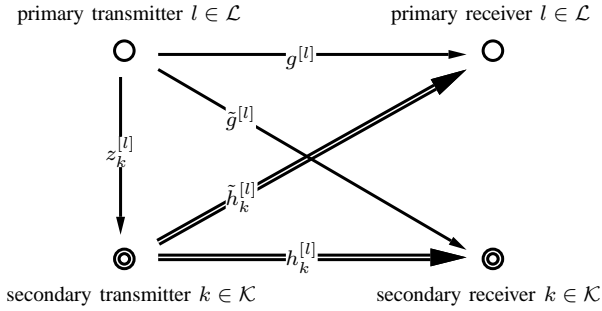

Consider a set $\setK = \br{1,\ldots,K}$ of secondary transmitters-receiver pairs and a set of orthogonal channels $\setL = \br{1,\ldots,L}$ licensed to primary users. Each secondary user (SU) wants to use a set of channels for communication. The system model is illustrated in \figurename~\ref{fig:sysmodel}. 

We assume that the distributed assignment of the channels to the SU can be done either using a coordinator or through direct communication between the SUs. In the stable matching model studied in Section \ref{sec:stable_matching}, we assume the existence of a coordinator which is connected to the SUs through low-rate links. In Section \ref{sec:walras_equilibrium}, we do not assume the existence of a coordinator, but require that the SUs can directly communicate with each other.

\subsection{Secondary System Performance}\label{sec:sys_model_su_performance}
An SU is allowed to access a set of channels if these are detected to be idle. We assume a primary user (PU) operates in a time-slotted fashion and starts transmission at the beginning and for the duration of a time-slot. Each SU at the beginning of the time slot is assumed to make a number $N$ of sensing observations in each channel $l$. The sensing problem of SU $k$ is the decision between two hypothesis on whether PU $l$ is active ($H_{k,1}^\ebr{l}$) or not ($H_{k,0}^\ebr{l}$). The two hyposesis correspond to:
\begin{align}
H_{k,0}^\ebr{l}: ~& x_{l,k}(t) = w_k(t), \quad t = 1,\ldots,N,\\
H_{k,1}^\ebr{l}: ~& x_{l,k}(t) = \sqrt{P_l} z_{k}^\ebr{l} s_l(t) + w_k(t),\quad t = 1,\ldots,N,
\end{align}
\noindent where $s_l(n) \sim \mathcal{CN}(0,1)$ is the transmitted signal of PU $l$, $P_l$ is the average primary transmission power, $w_k(n) \sim \mathcal{CN}(0,\sigma^2)$ is additive white Gaussian noise, and $z_{k}^\ebr{l} \sim \mathcal{CN}(0,1)$ is the quasi-static block flat-fading channel from PU $l$ to SU $k$ assumed constant during the time-slot.

Let $f^\ebr{l}_{k} = \text{Pr}(H_{k,1}^\ebr{l} \mid H_{k,0}^\ebr{l})$ and $d^\ebr{l}_{k} = \text{Pr}({H_{k,1}^\ebr{l} \mid H_{k,1}^\ebr{l}})$ be the \emph{false alarm} and \emph{detection probability} of the detector at SU $k$, respectively. The access probability of SU $k$ in channel $l$ is given as $\theta^\ebr{l}_k = (1 - \vartheta^\ebr{l}) (1- f^\ebr{l}_k) + \vartheta^\ebr{l} (1- d^\ebr{l}_k)$, where $\vartheta^\ebr{l}$ is PU $l$ transmission probability.

After spectrum sensing, an SU can be assigned a channel $l$ if he detects that PU $l$ is idle. The signal from secondary transmitter $k$ received at secondary receiver $k$ on channel $l$ is %written as
\begin{equation}
y^\ebr{l}_k = \left\{
           \begin{array}{ll}
             h^\ebr{l}_{k} \sqrt{P_k} s_k + \tilde{g}^\ebr{l}_{k} \sqrt{P_l} s_l + w_k, & \hbox{PU $l$ is active;} \\
             h^\ebr{l}_{k} \sqrt{P_k} s_k + w_k, & \hbox{otherwise,}
           \end{array}
         \right.
\end{equation}
\noindent where $s_k\sim \mathcal{CN}(0,1)$ is the transmitted signal, $P_k$ is the transmission power assumed to be the same in all channels, $h^\ebr{l}_{k}$ is the channel from secondary transmitter $k$ to its receiver, and $w_k \sim \mathcal{CN}(0,\sigma^2)$ is additive white Gaussian noise. We assume that $P_k$ is fixed and equal for all SUs and define the signal-to-noise ratio (SNR), used in the simulations, as $\text{SNR} := P_k /\sigma^2$.

The average \emph{achievable rate} in bits/s/Hz of SU $k$ in channel $l$ can be formulated as
\begin{multline}\label{eq:rate1}
u^{\su}_k(l) = (1 - \vartheta^\ebr{l}) (1- f^\ebr{l}_k) \log_2\pp{1 +\frac{P_k \sabs{h^\ebr{l}_{k}}}{\sigma^2}}\\ + \vartheta^\ebr{l}(1- d^\ebr{l}_k) \log_2\pp{1 + \frac{P_k \sabs{h^\ebr{l}_{k}}}{\sigma^2 + P_l \sabs{\tilde{g}^\ebr{l}_{k}} }}.
\end{multline}
%\begin{multline}\label{eq:rate1}
%u^{\su}_k(l) = \underbrace{(1 - \vartheta^\ebr{l}) (1- f^\ebr{l}_k) \log_2\pp{1 +\frac{P_k \sabs{h^\ebr{l}_{k}}}{\sigma^2}}}_{\text{SU transmits alone}} \\ + \underbrace{\vartheta^\ebr{l}(1- d^\ebr{l}_k) \log_2\pp{1 + \frac{P_k \sabs{h^\ebr{l}_{k}}}{\sigma^2 + P_l \sabs{\tilde{g}^\ebr{l}_{k}} }}}_{\text{SU transmits with PU}}.
%\end{multline}
\noindent The first term in the summation above is the average achievable rate when the PU is idle (also called opportunistic rate \cite{Scutari2013}) and the second term corresponds to the achievable rate on transmission simultaneously with the PU.

If an SU $k$ is assigned the set $\setB \subseteq \setL$ of channels, his average sum-rate is then
\begin{equation}\label{eq:utility_secondary}
u^{\susum}_k(\setB) = \sum\nolimits_{l \in \setB} u^{\su}_{k}(l),
\end{equation}
\noindent where $u^{\su}_{k}(l)$ is defined in \eqref{eq:rate1} and $u^{\su}_k(\emptyset) = 0$. In this work, we introduce the following channel assignment constraint: The maximum number of channels an SU $k$ can be assigned to is restricted to a maximum of $q_k \in \mathbb{N}$, called $quota$, and is assumed to be fixed for each SU.%

\subsection{Primary System Performance}\label{sec:sys_model_pu_performance}
% primary utility
%The received signal at a primary receiver $l$ is written as
%\begin{equation}
%\tilde{y}^\ebr{l}_k = \left\{
%           \begin{array}{ll}
%             g^\ebr{l} \sqrt{P_l} s_l + \tilde{h}^\ebr{l}_{k} \sqrt{P_k} s_k + w_l, & \hbox{SU $k$ is active in $l$;} \\
%             g^\ebr{l}_{k} \sqrt{P_l} s_l + w_l, & \hbox{otherwise,}
%           \end{array}
%         \right.
%\end{equation}
%\noindent where $s_l\sim\mathcal{CN}(0,1)$ is the transmitted signal of primary transmitter $l$, $g^\ebr{l}$ is the channel between primary transmitter $l$ and primary receiver $l$, and $P_l$ is the used transmission power. The channel from secondary transmitter $k$ to primary receiver $l$ is $\tilde{h}^\ebr{l}_{k}$, and $w_l \sim \mathcal{CN}(0,\sigma^2)$ is additive white Gaussian noise.
%\begin{multline}\label{eq:utility_primary}
%u^{\pu}_{l}(k) = \vartheta^\ebr{l} d^\ebr{l}_k \log_2\pp{1 +\frac{P_l \sabs{g^\ebr{l}}}{\sigma^2}}\\
%\quad + \vartheta^\ebr{l}(1 - d^\ebr{l}_k) \log_2\pp{1 + \frac{P_l \sabs{g^\ebr{l}}}{\sigma^2 + P_k \sabs{\tilde{h}^\ebr{l}_{k}} }}.
%\end{multline}
%\begin{equation}\label{eq:utility_primary_phi}
%u^{\pu}_{l}(\emptyset) = \vartheta^\ebr{l} \log_2\pp{1 +{P_l \sabs{g^\ebr{l}}}/{\sigma^2}}.
%\end{equation}

If channel $l$ is assigned to SU $k$, then the performance of PU $l$ decreases in both \emph{probability of misdetection} $(1- d^\ebr{l}_k)$ and interference $P_k \sabs{\tilde{h}^\ebr{l}_{k}}$, where $\tilde{h}^\ebr{l}_{k}$ is the channel from secondary transmitter $k$ to primary receiver $l$. Accordingly, we formulate the utility function of a PU $l$ as:
\begin{equation}\label{eq:utility_primary}
u^{\pu}_{l}(k) = \phi_l(1-d^\ebr{l}_k, P_k \sabs{\tilde{h}^\ebr{l}_{k}}), \\
\end{equation}
\noindent where $\phi_l(x,y) \leq \phi_l(x',y')$ for $x \geq x'$ and $y \geq y'$. If no SU is active in channel $l$, the interference-free utility of PU $l$ is $u^{\pu}_{l}(\emptyset) = \phi_l(1, 0)$. We additionally define \emph{self-matching} of channel $l$ as
\begin{equation}\label{PU_selfmatching}
u^{\pu}_l(l) = \underline{u}^{\pu}_{l} \le u^{\pu}_l(\emptyset), \quad l \in \setL,
\end{equation}
\noindent where the value $\underline{u}^{\pu}_{l}$ reflects a threshold for a QoS requirement of PU $l$. This QoS requirement will be incorporated later in the stable matching framework in Section \ref{sec:stable_matching}.

Since the utility of a PU is largest without interference from SUs, the region
\begin{equation}\label{eq:rate_region_box_PU}
%\begin{split}
\mathcal{R}^\pu = \{(r_1,\ldots,r_L)\in \mathbb{R}_+^L \mid r_l \leq u^\pu_1(\emptyset), l \in \setL\},
%\end{split}
\end{equation}
\noindent contains all jointly achievable performances for the PUs. A subset of $\mathcal{R}^\pu$, specified as
\begin{multline}\label{eq:rate_region_PU_tilde}
%\begin{split}
\widetilde{\mathcal{R}}^\pu = \{(u^\pu_1(a_1),\ldots,u^\pu_L(a_L)) \in \mathcal{R}^\pu \mid a_l \in \setK \cup \{l\}, \\ a_l \neq \emptyset, \sum\limits_{\substack{a_{l'} = k \\ l' \in \setL}} 1 \leq q_k, k \in \setK , l \in \setL \},
%\end{split}
\end{multline}
\noindent does not contain the performance tuples in which a PU operates alone, i.e., $a_l \neq \emptyset$ for all $l$, but only the performance tuples of the PUs when SUs are assigned to them or when the PUs are self-matched as specified in \eqref{PU_selfmatching}. Also, the region in \eqref{eq:rate_region_PU_tilde} takes into account the quota restrictions on the SUs. Later, we utilize the definition of $\widetilde{\mathcal{R}}^\pu$ to relate to existing efficiency results for stable matching.

\section{Problem Description}\label{sec:problem_def}
%%%%%%%%%%%%%%%%%%%%%%%%%%%%%%%%%%%%%%
%
%
%%%%%%%%%%%%%%%%%%%%%%%%%%%%%%%%%%%%%%
Our objective is to find an assignment of primary channels to the SUs through distributed mechanisms. Define the \emph{assignment variable}
\begin{equation}
x(\setB,k) = \left\{
  \begin{array}{ll}
    1, & \hbox{$\setB \subseteq \setL$ is assigned to SU $k \in \setK$;} \\
    0, & \hbox{otherwise.}
  \end{array}
\right.
\end{equation}
%
% constraints
\noindent In addition, define the following set of assignment constraints:
\begin{align}\label{msr:constraint1}
x(\setB,k)  \in \br{0,1}, &~~ \forall \setB \subseteq \setL, k \in \setK,\tag{C1}\\ \label{msr:constraint2}
\sum\nolimits_{\setB \ni l} \sum\nolimits_{k \in \setK} x(\setB,k) \leq 1, &~~ \forall l \in \setL, \tag{C2}\\ \label{msr:constraint3}
\sum\nolimits_{\setB \subseteq \setL} x(\setB,k) \leq 1, &~~ \forall k \in \setK,\tag{C3}\\ \label{msr:constraint4}
\abs{\setB}  x(\setB,k) \leq q_k,&~~  \forall \setB \subseteq \setL, k \in \setK, \tag{C4}\\ \label{msr:constraint5}
u^\pu_l(k) x(\setB,k) \geq \underline{u}^\pu_{l} x(\setB,k),&~~  \forall l \in \setB, \forall \setB \subseteq \setL, k \in \setK. \tag{C5}
\end{align}
\noindent Constraint \eqref{msr:constraint2} ensures that only one SU is allocated per channel,\footnote{We impose the orthogonality constraint on the channel assignment because the frameworks we exploit from stable matching and competitive markets do not take \emph{externalities} \cite{Bando2012} into account which would exist in nonorthogonal assignments. In our context, externalities are the interdependencies of the allocation of channels to some users on a given channel assignment to a specific user. Settings with externalities are generally much more complex to analyze, especially regarding the stability of distributed resource allocation algorithms. Recent application of stable matching with externalities for user association in small cell networks can be found in \cite{Namvar2014}.} and constraint \eqref{msr:constraint3} ensures that each SU is associated with one subset of $\setL$. The user quota constraint is in \eqref{msr:constraint4} and constraint \eqref{msr:constraint5} specifies a QoS threshold for each PU.

We will utilize the definition of the constraints $\eqref{msr:constraint1} - \eqref{msr:constraint5}$ to describe the coordination and cooperation mechanisms we propose in this work.%
\subsection{Coordination Mechanism}\label{sec:problem_def_coor}%
%%%%%%%%%%%%%%%%%%%%%%%%%%%%%%%%%%
Our first objective is to propose a low complexity coordination algorithm which matches the SUs to the PU channels exploiting the existence of a coordinator. The SUs and PU channels form two agent sets and each agent has a preference over each agent in the other set. These preferences are according to the utility functions defined in Section \ref{sec:sys_model_su_performance} and Section \ref{sec:sys_model_pu_performance}, respectively. For this purpose, we use many-to-one stable matching (Section \ref{sec:stable_matching}) as an assignment of the channels to the SUs. 

Generally, distributed implementation of stable matching requires communication between one agent set and the other in order to exchange proposals. In this work, we assume that the coordinator receives proposals from the SUs and accepts or rejects them on behalf of the PU channels. In order to implement the coordinated stable matching algorithm, the information which should be available at SU $k$ is $u^{\su}_k(l)$ in \eqref{eq:rate1} for all $l\in \setL$ and his quota $q_k$, while the coordinator needs the information of $u^{\pu}_{l}(k)$ in \eqref{eq:utility_primary} for all $l\in \setL$ and all $k\in \setK$. The information $u^{\pu}_{l}(k) = \phi_l(1-d^\ebr{l}_k, P_k \sabs{\tilde{h}^\ebr{l}_{k}})$ at SU $k$ requires the knowledge of the probability of misdetection $1-d^\ebr{l}_k$, which is known at the SU, and the interference at the primary receiver $l$. Assuming time division duplex (TDD) systems, the channel gain $\sabs{\tilde{h}^\ebr{l}_{k}}$ from secondary transmitter $k$ to primary receiver $l$ is almost identical to the channel gain from primary receiver $l$ to secondary transmitter $k$ and can be made available at the SU during channel estimation to calculate the interference $P_k \sabs{\tilde{h}^\ebr{l}_{k}}$. Having this information, we assume that each SU $k$ forwards $u^{\pu}_{l}(k)$ for all $l\in \setL$ to the coordinator in an initialization phase. 

If the coordinator also knows the utilities of the SUs, $u^{\su}_k(l)$ for all $k\in \setK$ and $l\in \setL$, then the Hungarian method\footnote{The Hungarian method \cite{Kuhn1955} is an algorithm that solves the fundamental one-to-one assignment problem in combinatorial optimization. The many-to-one channel assignment with quotas in \eqref{eq:optimization2} is a form of the generalized assignment problem for which the Hungarian method can be applied when $q_k$-many virtual SUs with a quota of one are introduced for each SU $k \in \setK$.} can be applied at the coordinator to find an assignment which satisfies $\eqref{msr:constraint1} - \eqref{msr:constraint5}$. The advantage of stable matching, however, is its flexibility to adapt to network dynamics and also to complexity requirements. We discuss these issues in Section \ref{sec:dist_sm}.

\subsection{Cooperative Mechanism}\label{sec:problem_def_coop}%
%%%%%%%%%%%%%%%%%%%%%%%%%%%%%%%%%%
The cooperative mechanism relies on direct communication between the SUs and does not require the existence of a coordinator as in the coordination mechanism. We consider the optimization of both the secondary and primary performance measures which is a multi-objective optimization problem. One method for solving multi-objective optimization problems is by optimizing the weighted sum of the objectives \cite{Marler:2004aa} which can be formulated for user $k$ utilizing resource set $\setB$ as:
\begin{equation}\label{eq:weightedsumutility}
W(\setB,k) = \lambda u^\susum_k(\setB) + (1 - \lambda) \sum\nolimits_{l \in \setB} u^\pu_l(k),
\end{equation}
\noindent with $\lambda \in [0,1]$. Here, $\lambda$ is a parameter which can be used to increase the priority of one objective to the other. If $\lambda$ is close to one, the secondary system performance is given more importance in the optimization above than the primary user performance, while if $\lambda$ is close to zero, the primary system performance is prioritized. The value of $\lambda$ must be defined in regard of the network specifications.

The integer optimization problem we are interested to solve is stated as follows:
\begin{equation}\label{eq:optimization2}
\maximize ~~ \sum\nolimits_{k \in \setK} \sum\nolimits_{\setB \subseteq \setL}  W(\setB,k)  x(\setB,k) \quad \text{s.t.}  ~~ \eqref{msr:constraint1} - \eqref{msr:constraint4}.
\end{equation}
The solution of \eqref{eq:optimization2} is a Walrasian equilibrium of an associated competitive market with indivisible goods studied in Section \ref{sec:walras_equilibrium}. The Walrasian equilibrium can be reached through a distributed English auction which we exploit to provide a decentralized and optimal cooperative channel assignment mechanism. 

In order to implement the cooperative mechanism to solve \eqref{eq:optimization2}, each SU $k$ must know $W(\{l\},k)$ for all $l \in \setL$, i.e., SU $k$ must know $u^{\pu}_{l}(k)$ in \eqref{eq:utility_primary} for all $l\in \setL$, his utility function $u^{\su}_k(l)$ in \eqref{eq:rate1} for all $l\in \setL$, the weight $\lambda$, and also his quota $q_k$. Moreover, all SUs must have knowledge of a common parameter $\alpha>0$ which will be used as a price incrementing factor.
\section{Many-to-One Stable Matching}\label{sec:stable_matching}%
We propose assigning SUs to the channels associated with the PUs by a framework for which stability serves as solution concept instead of optimality.
In cognitive radios, where SU access on channels is opportunistic and complex regulation entities are commonly absent, a stable distributed assignment process is favored. The applied framework involves a two-sided matching market where a coordinator is acting on behalf of the PU side to support the decision-making. We assume that one primary channel $l \in \mathcal{L}$ is matched to one SU $k \in \mathcal{K}$ while the latter can be assigned to up to $q_k$ primary channels, where $q_k \in \mathbb{N}$ is the maximum \textit{matching quota}, see Section \ref{sec:sys_model}. This resource allocation is named college admission \cite{Gale1962} or hospitals/residents problem \cite{Omalley2007} in the stable matching literature.

A stable matching is produced by a distributed process that matches together preference relations of the primaries and the secondaries over the other agents each. The order of preferences is given by the strictly ranked rate utilities in \eqref{eq:rate1} and \eqref{eq:utility_primary}.
In some cases, matching one agent with himself (denoted as being unmatched) might be preferred to a matching with other agents in the stable matching framework. Therefore, we define the matching of SU $k \in \setK$ to himself as $u^{\su}_k(k) = u^{\su}_l(\emptyset) = 0$ which means that the device does not transmit on any primary channel. On the other hand, matching a PU with himself should intuitively lead to the utility $u^{\pu}_{l}(\emptyset) = \phi_l(1, 0)$ as in Section~\ref{sec:sys_model_pu_performance}, since the PU occupies its channel $l$ alone and no SU transmits in $l$.
However, we specify the utility of a self-matched primary channel to a value $\underline{u}^{\pu}_{l}$, see \eqref{PU_selfmatching}, reflecting a threshold for a QoS requirement of the PU.
In doing so, a PU prefers being matched to a SU if its own utility remains higher than the threshold and prefers self-matching if matching with a SU does not guarantee QoS. The coordinator is in charge of monitoring the QoS issues of the PU in the stable matching framework.

\subsection{Stable Matching Model}
The stable matching problem is described by the tuple \cite{Jorswieck2011} $
\langle\setL,\setK, \{u_l^{\pu}\}_{l \in \setL}, \{u_k^{\su}\}_{k \in \setK}, \{q_k\}_{k\in\setK} \rangle,$ \noindent where $\setL$ is the set of primary channels, $\setK$ is the set of SUs, $u_l^{\pu}$ and $u_k^{\su}$ are the utility functions of the PUs and SUs given in \eqref{eq:utility_primary} and \eqref{eq:utility_secondary}. %respectively.
The quotas $q_k$ are associated with the SUs.
\begin{definition}\label{def:Matching}
A matching $M$ is from the set $\setK \cup \setL$ into the set of unordered family of elements of $\setK \cup \setL$ such that\footnote{Definition 1 is adopted from \cite{Roth1990} despite the fact that our framework does not fill an under-subscribed matching set with multiple copies of the self-matched agent.}
\begin{enumerate}
\item $\abs{M(l)} = 1$ for every PU $l \in \setL$ where $M(l) = l$ if $M(l) \notin \setK$,
\item $1 \le \abs{M(k)} \le q_k$ for every SU $k \in \setK$ where $M(k) = k$ if $M(k) \not\subset \setL$, %if the number of PUs in $M(k)$ say $r$ is less than $q_k$, then $M(k)$ contains $q_k - r$ copies of $k$
\item $M(l) = k$ if and only if $l \in M(k)$.
\end{enumerate}
\end{definition}
In Definition \ref{def:Matching}, $M(l)$ denotes the matched SU of PU $l$ or self-matching and $M(k)$ denotes the subset of PUs matched to SU $k$ or self-matching, respectively.

\begin{definition}\label{def:individually_rational}
The matching $M$ is \emph{individually rational} if there exists no PU $l \in \setL$ for which $u^\pu_l(l) > u^\pu_l(M(l))$ and no SU $k \in \setK$ for which $u^\su_k(k) > u^\su_k(j)$, $j \in M(k)$ \cite{Roth1990}.
\end{definition}
Individually rational matching ensures that no user, primary or secondary, would prefer being matched to himself than with the current matching.

\begin{definition}
The matching $M$ is \emph{blocked} by the pair $(k,l) \in \mathcal{K} \times \mathcal{L}$ if (i) $u^\pu_l(k) > u^\pu_l(M(l))$ \textbf{and} (ii) $|M(k)| < q_k$ and $u^\su_k(l) > 0$ \textbf{or} $u^\su_k(l) > u^\su_k(l')$ for some $l' \in M(k)$.\end{definition}
Accordingly, a matching is blocked by $(k,l)$ if these prefer each other to their current matching.

\begin{definition}
A matching $M$ is \emph{stable} if it is individually rational and not blocked by any pair $(k,l) \in \mathcal{K} \times \mathcal{L}$.
\end{definition}

There may exist several stable matchings. Let all stable matchings lead respectively to the SU and PU performance regions $\mathcal{R}^\su_{SM}$ and $\mathcal{R}^\pu_{SM} \subseteq \widetilde{\mathcal{R}}^\pu$, where $\widetilde{\mathcal{R}}^\pu$ is in \eqref{eq:rate_region_PU_tilde}. Next, we provide an algorithm which reaches a stable matching and reveal its performance in these regions.

\subsection{Distributed Implementation of Stable Matching}\label{sec:dist_sm}
Algorithm \ref{alg:sm_su_optimal} implements a distributed coordination mechanism proposed in \cite{Roth1984} to deliver a stable matching. Here, the SUs start proposing to be matched with their preferred PU channels (Line 2) and a low-complex coordinator responds on behalf of the PUs (starting Line 3). The information needed at the SUs and the coordinator prior to the execution of Algorithm \ref{alg:sm_su_optimal} is stated in Section \ref{sec:problem_def_coor}. In the given protocol, SU $k$ proposes to be matched to a channel $l$ by sending a message $\Psi_{k}$ to the coordinator if $k$ has not reached its quota and prefers this channel it is not already matched with (condition in Line 1). The message $\Psi_k$ can be of $\lceil \log_2 (l) \rceil$ bits which is the length of the base-2 equivalent of $l$. The coordinator reacts to the proposal by sending a one bit message to an SU $k$ to indicate acceptance or rejection. The coordinator accepts SU $k$ on a channel $l$ (Line 7 and 8) only if the QoS requirement of the PU is fulfilled (Line 4). Otherwise, the coordinator rejects SU $k$ (Lines 5 and 6). Also, in order to reduce the total number of iterations of the algorithm, the coordinator triggers messages to exclude selected SUs from proposing to certain channels (Lines 9 and 10). Including the rejection information in Line 6, these messages are of $L$ bits and indicate for each SU which channels he need not propose to.

% complexity
In Algorithm \ref{alg:sm_su_optimal}, each SU proposes at most \emph{once} to be matched with a specific resource. Thus, the worst case total number of proposals by an SU to the coordinator is $L$. 
\begin{proposition}
The worst case number of bits that is exchanged between one SU and the coordinator during Algorithm \ref{alg:sm_su_optimal} is $L^2 + L+ \sum_{l=1}^L \lceil \log_2 (l) \rceil$. 
\end{proposition}
\begin{IEEEproof}
The term $\sum_{l=1}^L \lceil \log_2 (l) \rceil$ is the total number of bits needed to indicate the channel indexes in the $L$ proposals from the SU. $L$ bits are needed in total to indicate the acceptance or rejection from the coordinator to the $L$ proposals from the SU. The term $L^2$ is due to $L$ bit messages sent from the coordinator to the SUs in Lines 9 and 10 of Algorithm \ref{alg:sm_su_optimal}.
\end{IEEEproof}
The actual number of proposals by a single SU depends on his quota and also on the matching of the channels to the other SUs. If the quota of an SUs is small, then a few proposals can be sufficient to reach the matching quota limit and stop the SU from further proposals. Also, when several SUs are already accepted on some channels, the number of channels which an SU can propose to may decrease due to Lines 9 and 10 in Algorithm \ref{alg:sm_su_optimal}. In Section \ref{sec:simulations}, we provide extensive simulations on the average number of proposals from an SU.

% In comparison, Hungarian method has higher complexity of $\mathcal{O}(n^3)$, with $n = \max\{\sum_{k\in\setK}q_k,L\}$. 
Observe in the implementation of stable matching that if new secondary users arrive to the network and propose to the coordinator to be matched to a set of channels, the coordinator can use Algorithm \ref{alg:sm_su_optimal} with the initialization of the current stable matching. In contrast, the application of the Hungarian method necessitates the network wide optimization problem to be solved again. Nevertheless, the stable matching algorithm can be terminated at any time instance associated to a desirable complexity level to retrieve an orthogonal matching of resources to the SUs. Such properties of the algorithm make it adaptable to changes in the network and also to specified complexity or implementation requirements. 

Initializing Algorithm \ref{alg:sm_su_optimal} with unmatched SUs and channels as $M(l) = \{ l \} \, \forall \, l$, $M(k) = \{ k \} \, \forall \, k$, the terminating state is an SU-optimal stable matching which is weak Pareto optimal\footnote{The set of all weak Pareto optimal points in a performance region $\mathcal{R}$ are defined as \cite[p. 14]{Peters1992} $\mathcal{W}(\mathcal{R})=\br{\mat{x} \in \mathcal{R} \mid \text{there is no } \mat{y} \in \mathcal{R} \text{{ with }} \mat{y} > \mat{x}}.$} in the set $\mathcal{R}^\su_{SM}$ according to \cite[Corollary 5.9]{Roth1990} but is the worst stable matching for the PUs \cite[Corollary 5.30]{Roth1990}. Next, we provide conditions under which our stable matching is unique and also sum-performance optimal for the PUs.
% uniqueness
\begin{theorem}\label{thm:uniqueSM}
For $q_k\geq L$ for all $k$, the stable matching is \emph{unique} and leads to the maximum sum performance point in $\widetilde{\mathcal{R}}^\pu$ defined in \eqref{eq:rate_region_PU_tilde}.
\end{theorem}
\begin{IEEEproof}
The proof is provided in Appendix \ref{proof:uniqueSM}.
\end{IEEEproof}
The result above generalizes the uniqueness result of one-to-one matching in \cite[Proposition III.1]{Yaffe2010} to the case of many-to-one matching.

%% SU-proposing algorithm
\begin{algorithm}[t!]
\begin{algorithmic}[1]

	%\SetKwInput{KwData}{Initialize}

	%\KwIn{matching market $\langle\setL,\setK, \{u_l^{\pu}\}_{l \in \setL}, \{u_k^{\su}\}_{k \in \setK}, \{q_k\}_{k\in\setK} \rangle$}
	%\KwData{$M(l) = \{ l \} \, \forall \, l$; $M(k) = \{ k \} \, \forall \, k$}		
		
	\While{some SU $k \in \setK$ is under-subscribed ($|M(k)|<q_k$ or $M(k)=k$) \textbf{and} $\max_{l \in \setL, l \notin M(k)} u^\su_k(l)> u^{\su}_k(k)$}

    \State \parbox[t]{\dimexpr\linewidth-\algorithmicindent}{\underline{Proposal by SU $k$:}
		send out index $l^{\star} = \arg\!\max_{l \in \setL, l \notin M(k)} u^\su_k(l)$ of most preferred PU\strut}		
    \State \underline{Coordinator Response:}

	\If{QoS is ensured, i.e. $u^\pu_{l^{\star}}(k) > \underline{u}^\pu_{l^{\star}}$}	
        \If{PU $l^{\star}$ is engaged to any SU $k^{\star} \neq k$}
		      \State \parbox[t]{\dimexpr\linewidth-\algorithmicindent-\algorithmicindent-\algorithmicindent}{inform $k^\star$ on releasing engagement with $l^\star$, giving $M(k^{\star}) = \{ k^{\star} \}$ if it was $|M(k^{\star})| = 1$ and $M(k^{\star}) = M(k^{\star}) \setminus \{ l^{\star} \}$ otherwise\strut}
%			\If{$|M(k^{\star})| = 1$}
%				\State $M(k^{\star}) = \{ k^{\star} \}$
%			\Else
%				\State $M(k^{\star}) = M(k^{\star}) \setminus \{ l^{\star} \}$
%			\EndIf
        \EndIf	
        \State accept engagement temporarily; set $M(l^{\star}) = \{ k \}$
        \State \parbox[t]{\dimexpr\linewidth-\algorithmicindent-\algorithmicindent}{inform SU $k$ on approving the engagement, giving $M(k) = \{ l^{\star} \}$ if $k$ was unmatched ($M(k) = \{ k \}$) and $M(k) = M(k) \cup \{ l^{\star} \}$ otherwise\strut}
	\EndIf		
	\ForAll{$i \in \setK$ such that $u^\pu_{l^{\star}}(i) < \max \{ u^\pu_{l^{\star}}(k) , \underline{u}^\pu_{l^{\star}} \} $}
	   \State \parbox[t]{\dimexpr\linewidth-\algorithmicindent-\algorithmicindent}{eliminate preference on $i$ ($u^\pu_{l^{\star}}(i) = 0$) and disqualify SU $i$ from proposing to $l^{\star}$ ($u^\su_{i}(l^{\star}) = 0$)\strut} % for bilateral preference elimination
	   %\State \parbox[t]{\dimexpr\linewidth-\algorithmicindent-\algorithmicindent}{eliminate preference on $i$ ($u^\pu_{l^{\star}}(i) = 0$) and send message to SU $i$ to ensure his preference elimination on $l^{\star}$ ($u^\su_{i}(l^{\star}) = 0$)\strut} % for bilateral preference elimination

    \EndFor
	\EndWhile			
	
\end{algorithmic}
\caption{\label{alg:sm_su_optimal}Distributed SU-proposing stable matching.} % \cite[Algorithm 5]{Omalley2007}
\end{algorithm}%
\section{Walrasian Equilibrium}\label{sec:walras_equilibrium}
In the previous section, we studied two-sided matching where the SUs on one side are matched to primary channels on the other. In this section, we study a market model where only one entity is represented (SUs) but its utility is the weighted combination of the utilities of the secondary and primary users given in \eqref{eq:weightedsumutility} in Section \ref{sec:problem_def}. Contrary to the previous section, we now do not assume the existence of a coordinator. However, we require that the SUs are able to communicate with each other in order to find an assignment of the channels which solves Problem \eqref{eq:optimization2}. The mechanism exploits the market model studied next.

\subsection{Competitive Market Model}
A competitive market with indivisible goods \cite{Gul1999} is composed of a set of consumers and a set of goods. The consumers in our setting are the SUs in $\setK$ and the goods correspond to the primary channels in $\setL$. Here, we define the \emph{unit-less}\footnote{The utility function can be made unit-less by dividing the terms with their associated unit of measure.} utility function of consumer $k$ using the utility function in \eqref{eq:weightedsumutility} with an additional restriction on his quota $q_k$ as follows
\begin{equation}\label{eq:utility_quota}
\begin{split}
U_k(\setA) = \maximize\limits_{~\setB \subseteq \setA} & ~~\lambda u^\susum_k(\setB) + (1 - \lambda) \sum\nolimits_{l \in \setB} u^\pu_l(k)\\ s.t. & ~~ \abs{\setB} \leq q_k.
\end{split}
\end{equation}
\noindent The function above is called the $q_k$-\emph{satiation} of the weighted sum-perfromance \cite[Section 2]{Gul1999}.

Each good $l$ has a price $p_l \geq 0$ which is in \emph{monetary units} and we assume that each SU is endowed with sufficient amount of monetary budget which enables him to buy bundles of goods. The unit-less \emph{net utility} of SU $k$ is
\begin{equation}\label{eq:net_utility}
v_k(\setB, \mat{p}) := U_k(\setB) - \sum\nolimits_{l \in \setB} p_l.
\end{equation}
\noindent Given the prices of the goods (primary channels) $\mat{p}=(p_1,\ldots,p_L)$, the \emph{demand correspondence} of SU $k$ is the set of goods which maximizes his net utility:
%\begin{equation}\label{eq:demand}
%\setD_k(\mat{p}) = \maximize_{\setB \subseteq \setL} \quad U_k(\setB) - \sum\nolimits_{l \in \setB} p_l.
%\end{equation}
\begin{equation}\label{eq:demand}
\setD_k(\mat{p})=\br{\setA \subseteq \setL \mid v_k(\setA,\mat{p}) \geq  v_k(\setB,\mat{p})~ \forall \setB \subseteq \setL}.
\end{equation}
Later in Algorithm~\ref{alg:demand} in Section~\ref{sec:Auction}, we specify a method to efficiently calculate the demand for each SU. The outcome of a competitive market is a Walrasian equilibrium which specifies the prices of the channels at which each SU buys the channels he demands and no channel is bought by more than one SU.
\begin{definition}\label{def:Walras}{\cite[Section 2]{Gul1999}}
A \emph{Walrasian equilibrium} is a tuple $(\mat{p},\setX_0,\setX_1,\ldots,\setX_K)$, where $\mat{p}\in \mathbb{R}_+^L$ is a price vector, and $(\setX_0,\ldots,\setX_K)$ is a partition of $\setL$, i.e., $\setX_k \cap \setX_j = \emptyset$ for all $k\neq j$, and $\bigcup_{k=0}^K \setX_k = \setL$, such that (i) for each $k \in \setK$, $v_k(\setX_k,\mat{p}) \geq v_k(\setA,\mat{p})$ for all $\setA \subseteq \setL$, and (ii) the price of any object in $\setX_0$ is zero.
\end{definition}

\noindent A Walrasian equilibrium \emph{exists} if and only if the utility function $U_k$ in \eqref{eq:utility_quota} satisfies \cite{Gul1999}:
\begin{enumerate}
\item \emph{monotonicity}: for all $\setA \subset \setB \subset \setL, U_k(\setA) \leq U_k(\setB)$,
\item \emph{gross substitutes condition}: for any two price vectors $\mat{p}'$ and $\mat{p}$ such that $\mat{p}' \geq \mat{p}$ (the inequality is componentwise), and any $\setA \in \setD_k(\mat{p})$, there exists $\setB \in \setD_k(\mat{p}')$ such that $\br{i \in \setA \mid p'_i= p_i} \subseteq \setB$.
\end{enumerate}
The gross substitute condition implies that if an SU demands a set of channels, and prices of some channels increase, the SU would still demand the channels whose prices did not change. %The gross substitute condition is proven in \cite[Theorem 3]{Kelso1982} to be sufficient for the existence of a Walrasian equilibrium. In \cite[Theorem 2]{Gul1999}, it is proven that this condition is also necessary.
\begin{theorem}\label{thm:existence}
A Walrasian equilibrium exists in our setting.
\end{theorem}
\begin{IEEEproof}
The proof is provided in Appendix \ref{proof:existence}.
\end{IEEEproof}

There is a direct relation between the solution of \eqref{eq:optimization2} and the Walrasian equilibrium of the associated competitive market model with indivisible goods \cite{Blumrosen2007}. The \emph{existence} of a Walrasian equilibrium ensures that the solution of \eqref{eq:optimization2} is identical to the solution of its linear programming relaxation \cite[Theorem 1.13]{Blumrosen2007} in which the integer constraint \eqref{msr:constraint1} is replaced by the convex constraint $x(\setB,k) \geq 0, \forall \setB \subseteq \setL, k \in \setK$. %Following Theorem \ref{thm:existence}, we have that the Walrasian equilibrium is the solution of Problem \eqref{eq:optimization2}.

Next, we will describe the English auction which is the price adjustment mechanism needed to reach the Walrasian equilibrium.%
\subsection{English Auction}\label{sec:Auction}
% --------------------------------------------------------------------------------------------------
%Auction algorithms for linear optimization problems \cite{Bertsekas1991}
%In each iteration, the base station sends the prices vector $\mat{p}$ to the users. Each user calculates his demand in \eqref{eq:demand} and sends it to the base station. The base station increases the prices of the resources in \emph{excess demand} and sends the updated prices to the users. This mechanism is repeated until the Walrasian equilibrium is reached at which no resources are in excess demand.

% consumer demand
%%%%%%%%%%%%%%%%%%%%%%%%%%%%%%%%%%%%%%%%%%%%%%%%%%%%%%%%%%%%%%%%%%
In the English auction \cite[Section 5]{Gul2000}, if a channel is simultaneously demanded by more than one SU then its price is increased. The auction terminates when each channel is demanded by at most one SU. This auction mechanism is within the combinatorial auctions frameworks classified in \cite{Zhang2013} and has been rarely applied in the context of wireless communication due to their complexity. In the following, we show that the steps required during the English auction in our model can be calculated efficiently.

In order to perform the English auction, we first need to efficiently calculate the consumer demand in \eqref{eq:demand}. Afterwards, we need to calculate the set of channels which are simultaneously demanded by more than one SU. This set is called the aggregate excess demand. These issues are addressed in the same order next.

The consumer demand in \eqref{eq:demand} seems at first sight hard to solve since a search over all $2^L$ subsets of $\setL$ is needed. Note that in \cite{Gul2000} no method is provided to calculate the demand, but is only assumed that the demand can be calculated efficiently. This assumption is known under the \emph{existence of a demand oracle}. We show that the consumer demand in our case can be solved in polynomial time with the number of channels using a greedy approach. First, we need the following result.%
\begin{lemma}\label{lem: quotas_not_violated}
If $p_l>0$ for all $l \in \setL$, then $\setA_k \in \setD_k(\mat{p})$ satisfies $\abs{\setA_k} \leq q_k$ for all $k \in \setK$.
\end{lemma}
\begin{IEEEproof}
The proof is provided in Appendix \ref{proof: quotas_not_violated}.
\end{IEEEproof}

From Lemma \ref{lem: quotas_not_violated}, if the prices of each good are strictly larger than zero, then a user demands at most as many resources as his quota. In order to calculate the demand of an SU, we use the following assumption to ensure that the best $q_k$ channels of a user $k$ are unique.%\footnote{If the assumption is violated, then the consumer demand may not be unique and hence the consumer is required to report multiple demand sets to the other SUs.}

\begin{assumption}\label{ass:unique_demand}
For any price vector $\mat{p}>0$, and for all consumers $k$, the net utilities satisfy $v_k(\{l\}, \mat{p}) \neq v_k(\{l'\}, \mat{p})$ for any two goods $l, l' \in \setL$.
\end{assumption}

%%%%%%%%%%%%%%%%%%%%%%%%%%%%%%%%%%%%%%%%%%%%%%%%%%%%%%%%%%%%%%%%%%%%%%%%%%%%%%%%%%%%%%%%%%%%%
\begin{algorithm}[t]
\caption{\label{alg:demand}Calculate demand $\setA_k$ of SU $k$.}
\begin{algorithmic}[1]
\State \textbf{Input}: {prices $\mat{p} = (p_1,\ldots,p_L)$; quota $q_k$}
\State \textbf{Init}: $\Pi = \{\pi_{1},\ldots, \pi_{L}\}$ with $\pi_{l} = v_k(\{l\}, \mat{p})$; $\setA_k=\emptyset$\;
\State sort $\Pi$ in descending order to obtain $\Pi^\text{sorted}$\;
\State set $\setA_k$ as the first $q_k$ elements in $\Pi^\text{sorted}$ which are strictly larger than zero.\;
%\State \textbf{Output}: {Demand set $\setA_k$.}
\end{algorithmic}
\end{algorithm}
%%%%%%%%%%%%%%%%%%%%%%%%%%%%%%%%%%%%%%%%%%%%%%%%%%%%%%%%%%%%%%%%%%%%%%%%%%%%%%%%%%%%%%%%%%%%%

\begin{theorem}\label{thm:greedy_demand}
For given prices $\mat{p}>0$, Algorithm~\ref{alg:demand} finds the consumer demand set which is the smallest subset of all sets in $\setD_k(\mat{p})$ defined in \eqref{eq:demand}.
\end{theorem}
\begin{IEEEproof}
The proof is provided in Appendix \ref{proof:greedy_demand}.
\end{IEEEproof}

% complexity of consumer demand
The complexity of calculating the demand of SU $k$ in Algorithm~\ref{alg:demand} requires a sorting algorithm such as Quick Sort which requires on average $\text{O}(L \log L)$ comparisons.
%%%%%%%%%%%%%%%%%%%%%%%%% excess demand

Later, in the distributed implementation of the Walrasian equilibrium in Section \ref{sec:dist_Walras}, it is required that each SU reports his demand set to the other SUs. If an SU knows all other SUs' demands, he can calculate the excess demand set which is composed of the channels simultaneously demanded by more than one SU. First, we need the following definitions before defining the excess demand set.

The \emph{requirement function} of consumer $k$ is defined as%\footnote{The requirement function is a dual rank function in matroid theory.}
\begin{equation}
K_k(\setB,\mat{p}) := \min\nolimits_{\setA \in D_k(\mat{p})} \abs{\setA \cap\setB},
\end{equation}
\noindent and reveals the smallest number of elements in common between $\setB$ and the demanded channels by SU $k$ at given prices $\mat{p}$. From \cite[Theorem 2]{Gul2000}, we have $K_k(\setB,\mat{p}) \leq K_k(\setB,\mat{q})$ for $\mat{p} \geq \mat{q}$ (componentwise inequality) and $p_l = q_l$ for $l \in \setL \setminus \setB$. This means that $K_k(\setB,\mat{p})$ decreases when the prices of the objects inside $\setB$ increase.

%%%%%%%%%%%%%%%%%%%%%%%%%%%%%%%%%%%%%%%%%%%%%%%%%%%%%%%%%%%%%%%%%%%%%%%%%%%%%%%%%%%%%%%%%%%%%
\begin{algorithm}[t]
\caption{\label{alg:excess_demand}Calculate excess demand $\setZ$.}
\begin{algorithmic}[1]
\State \textbf{Input}: {demand $\setA_1,\ldots,\setA_K$}
\State \textbf{Init}: $\setZ=\emptyset$
\For{$l =1,\ldots, L$}
    \ForAll{$k,j =1,\ldots, K$ \textbf{and} $j \neq k$}
        %\For{$j =1,\ldots, K$}
            \If{$l \in \setA_k \cap \setA_j$}
                \State $\setZ = \setZ \cup \{l\}$
            \EndIf
        %\EndFor
    \EndFor
\EndFor
%\State \textbf{Output}: {Excess demand $\setZ$.}
\end{algorithmic}
\end{algorithm}%
%%%%%%%%%%%%%%%%%%%%%%%%%%%%%%%%%%%%%%%%%%%%%%%%%%%%%%%%%%%%%%%%%%%%%%%%%%%%%%%%%%%%%%%%%%%%%

Since the demand from Algorithm~\ref{alg:demand} is the smallest subset of all demand sets following Theorem \ref{thm:greedy_demand}, we have the following result:
\begin{corollary}
The requirement function can be calculated as $K_k(\setB,\mat{p}) = \abs{\setA_k(\mat{p}) \cap\setB}$, where $\setA_k(\mat{p})$ is the demand set calculated using Algorithm~\ref{alg:demand}.
\end{corollary}

Define the function which counts the number of times each channel in $\setB$ is demanded as
\begin{equation}
K_{\setK}(\setB,\mat{p}) := \sum\nolimits_{k=1}^K K_k(\setB,\mat{p}).
\end{equation}
\noindent From \cite[Corollary of Theorem 3]{Gul2000}, a necessary condition that any channel in set $\setB$ is not demanded by more than one SU at the same time is $K_{\setK}(\setB,\mat{p}) - \abs{\setB} \leq 0$. Hence, in Walrasian equilibrium with prices $\mat{p}*$, it must hold $K_{\setK}(\setB,\mat{p}*) - \abs{\setB} \leq 0$ for all $\setB \subseteq \setL$. Define
\begin{multline}\label{eq:max_dem}
\mathcal{O}(\mat{p}) := \{\setA \subseteq \setL \mid K_{\setK} (\setA,\mat{p}) - \abs{\setA} \geq K_{\setK} (\setB,\mat{p}) - \abs{\setB},\\ \text{ for all } \setB \subseteq \setL\},
\end{multline}
\noindent which collects the set of channels $\setB \in \mathcal{O}(\mat{p})$ that maximize $K_{\setK}(\setB,\mat{p}) - \abs{\setB}$. The \emph{excess demand set} $\mathcal{Z}(\mat{p})$ is the smallest element of $\mathcal{O}(\mat{p})$ and can be calculated using Algorithm~\ref{alg:excess_demand} by checking whether each channel is simultaneously demanded by more than one SU. Algorithm~\ref{alg:excess_demand} requires in worst case $L K$ calculations.

\begin{theorem}\label{thm:excess_demand}
For given prices $\mat{p}$, Algorithm~\ref{alg:excess_demand} finds the excess demand set $\setZ(\mat{p})$.
\end{theorem}
\begin{IEEEproof}
The proof is provided in Appendix \ref{proof:excess_demand}.
\end{IEEEproof}

\subsection{Distributed Implementation of Walrasian Equilibrium}\label{sec:dist_Walras}

The English auction, proposed in \cite{Gul2000} and proven to reach a Walrasian equilibrium, can be implemented by an auctioneer (coordinator) which, upon collecting the demands from all the users, updates the prices and broadcasts them to the SUs. However, since we do not assume the existence of a coordinator in this section, we formulate a cooperative implementation based on the SUs exchanging the channel demands between themselves. Instead of updating the prices at the auctioneer, each SU can update the prices locally knowing the demands of all SUs. In Algorithm~\ref{alg:EnglishAuction}, we provide an implementation of this mechanism. As in the stable matching coordination in Section \ref{sec:stable_matching}, the SUs need only communicate the indices of the channels they demand. Thus, given $L$ channels, each SU needs to send $L$ bits of information to the other SUs to reveal his demand. Specifically, SU $k$ sends the $L$ bit message $\Psi_{k}$ to the other SUs with $[\Psi_k]_l = 1$ if $l \in \setA_k$ and $[\Psi_{k}]_l = 0$ otherwise. Given the demand sets from all SUs, each SU calculates the excess demand set and updates the prices by incrementing the prices of the channels in excess demand by a factor $\alpha$. Note that only in case the demand set of an SU has changed it is necessary that the SU broadcasts this update to the other SUs.

The choice of the price incrementing factor $\alpha$ for the prices influences the speed of convergence of the algorithm. For sufficiently small $\alpha$, i.e., $\alpha \rightarrow 0$, Algorithm~\ref{alg:EnglishAuction} converges to the Walrasian equilibrium (Definition \ref{def:Walras}). For relatively large $\alpha$, some channels may not be demanded by any SUs. The reason for this is that for a channel which has been demanded by more than one SU, the price update of this channel does not take into account the SUs' utilities such that a high price incrementing factor can make the channel suddenly unattractive to all SUs. Algorithm~\ref{alg:EnglishAuction} is guaranteed to converge since prices of the channels can only be incremented and the SU utilities are finite valued.

\begin{algorithm}[t]
\caption{\label{alg:EnglishAuction}Implementation of Walrasian Equilibrium by modified English Auction.}
\begin{algorithmic}[1]
\State \textbf{Input}: {price incrementing factor $\alpha > 0$}
\State \textbf{Init}: $p^{(0)}_l = \epsilon, l \in \setL$; $t=0$\;
\Repeat
\State \parbox[t]{\dimexpr\linewidth-\algorithmicindent}{Each SU $k$ calculates $\setA^{(t)}_k$ using Algorithm~\ref{alg:demand} and broadcasts it to all SUs\strut}\;
\State \parbox[t]{\dimexpr\linewidth-\algorithmicindent}{Each SU calculates $\setZ(\mat{p}^{(t)})$ using Algorithm~\ref{alg:excess_demand}\strut}\;
\State \parbox[t]{\dimexpr\linewidth-\algorithmicindent}{Each SU updates the prices as
\begin{equation}
\mat{p}^{(t+1)} = \mat{p}^{(t)} + \alpha \mat{\delta}(\mat{p}^{(t)}) 
\end{equation}
\begin{equation}\nonumber
\text{ with } \delta_i(\mat{p}^{(t)}) = \left\{
  \begin{array}{ll}
    1, & \hbox{$i \in \mathcal{Z}(\mat{p}^{(t)})$;} \\
    0, & \hbox{otherwise}
  \end{array}\right.
\end{equation}
}
\State $t = t + 1$\;
\Until{$\setZ(\mat{p}^{(t-1)}) = \emptyset$}
%\State \textbf{Output}: {Assignment $\setA^{(t-1)}_1,\ldots,\setA^{(t-1)}_K$.}
\end{algorithmic}
\end{algorithm}%

\section{Numerical Results}\label{sec:simulations}
\begin{figure}%[!ht]
\centering
    \subfloat[$q_k=1, \forall k\in\setK$ \label{fig:region_snr0_q1}]{\includegraphics[height=3.9cm,clip]{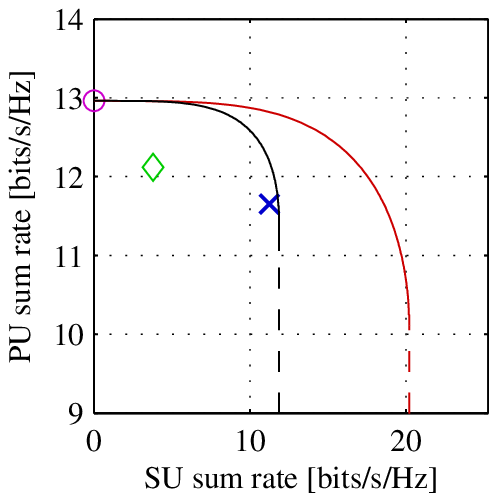}}
    \hspace{0.1cm}
    \subfloat[$q_k=2, \forall k\in\setK$ \label{fig:region_snr0_q2}]{\includegraphics[height=3.9cm,clip]{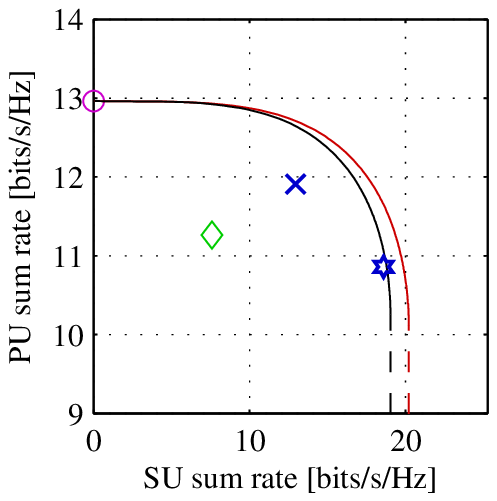}}
    \hspace{0.1cm}
    \subfloat[$q_k=L, \forall k\in\setK$ \label{fig:region_snr0_q20}]{\includegraphics[height=3.9cm,clip]{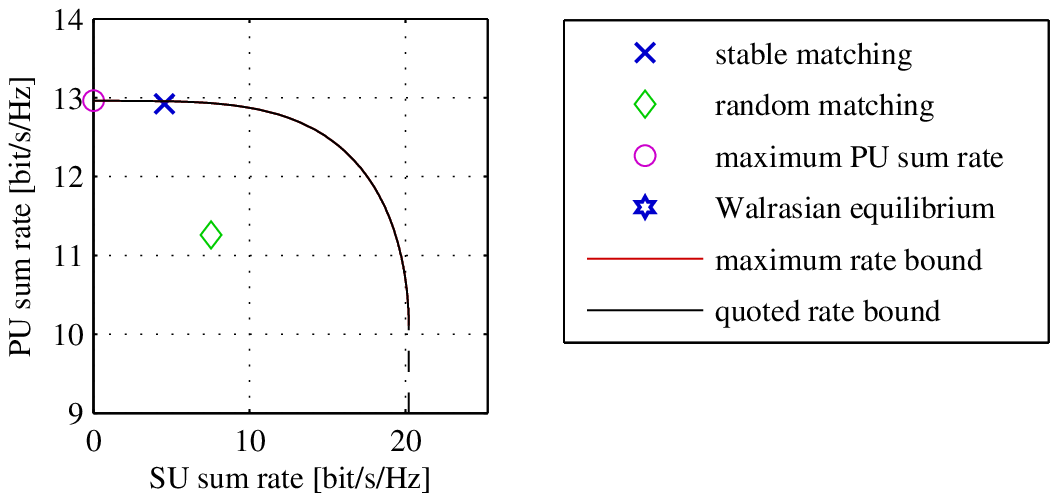}}
    %\hspace{0.5cm}
    %\subfloat[Legend]{\includegraphics[height=4cm,clip]{figures/region_0db_legend2.eps}}
    \caption{Comparison of stable matching and Walrasian equilibrium in the PU and SU sum rate regions for different quotas at 0 dB SNR.}
	\label{fig:region_snr0}
\end{figure}

%% general Parameters
We assume an energy detector is used at each SU $k$ with a detection threshold $\gamma^\ebr{l}_k$.
The {false alarm probability} and {probability of detection} of the energy detector are respectively approximated by $f^\ebr{l}_{k} = Q\pp{\frac{\gamma^\ebr{l}_k - N \sigma^2}{\sigma^2 \sqrt{2N}}} \text{ and } d^\ebr{l}_{k} = Q\pp{\frac{\gamma^\ebr{l}_k - N (\sigma^2 + P_l \sabs{z_{k}^\ebr{l}} )}{\sqrt{2N \sigma^2(\sigma^2 + 2 P_l \sabs{z_{k}^\ebr{l}})}}}$ \cite{Quan2008}.
We fix $f^\ebr{l}_{k}=0.05$ and choose $N = 20$ sensing observations. By calculating the threshold $\gamma^\ebr{l}_k$, we can determine the detection probability. The probability of primary transmission of all PUs is set to $\vartheta^\ebr{l} = 0.75$.
%\noindent and
%\begin{equation}\label{eq:user_detection}
%d^\ebr{l}_{k} = Q\pp{\frac{\gamma^\ebr{l}_k - N (\sigma^2 + P_l \sabs{z_{k}^\ebr{l}} )}{\sqrt{2N \sigma^2(\sigma^2 + 2 P_l \sabs{z_{k}^\ebr{l}})}}}.
%\end{equation}

For simulations, we adopt the primary utility function $\phi_l$ to be the average achievable rate:
\begin{multline}\label{eq:utility_primary_rate}
\phi_l(1-d^\ebr{l}_k, P_k \sabs{\tilde{h}^\ebr{l}_{k}}) = \vartheta^\ebr{l} d^\ebr{l}_k \log_2\pp{1 +\frac{P_l \sabs{g^\ebr{l}}}{\sigma^2}}\\ + \vartheta^\ebr{l}(1 - d^\ebr{l}_k) \log_2\pp{1 + \frac{P_l \sabs{g^\ebr{l}}}{\sigma^2 + P_k \sabs{\tilde{h}^\ebr{l}_{k}} }},
\end{multline}
and assume no specific QoS requirements of the PUs.

%%%%%%%%%%%%%%%%%%%%%%%%%%%%%%%%%%%%%%%%%%%%%%%%%%%%%%%%%%%%%
% performance region of stable matching outcome for different quota
%%%%%%%%%%%%%%%%%%%%%%%%%%%%%%%%%%%%%%%%%%%%%%%%%%%%%%%%%%%%%
In \figurename~\ref{fig:region_snr0}, we plot the average performance region of the primary (y-axis) and secondary systems (x-axis), where the number of SUs is $K=10$ and number of primary channels is $L=20$. All channels are independently and identically Rayleigh distributed and the simulations are averaged over $10^3$ random instances.

In \figurename~\ref{fig:region_snr0}, the region inside the quoted rate bound includes only the performance of the SUs and PUs in the channels in which the SUs are assigned to. This region is included in the region marked as maximum rate bound which includes all channel assignment possibilities to the SUs without quota restrictions. The quoted rate bound includes the stable matching and Walrasian equilibrium channel assignments. We generate both boundaries using the Hungarian optimization method \cite{Kuhn1955} which finds the optimal channel assignment by maximizing the weighted sum rate of SUs and PUs in a centralized way, see Section~\ref{sec:problem_def}.
%
%%%%%%%%%%%%%%%%%%%%%%%%%%%%%%%%%%%%%%%%%%%%%%%%%%%%%%%%%%%%%
% rate analysis of stable matching outcome vs. SNR
%%%%%%%%%%%%%%%%%%%%%%%%%%%%%%%%%%%%%%%%%%%%%%%%%%%%%%%%%%%%%
\begin{figure}[t]
	\centering
	\includegraphics[width=\linewidth,clip]{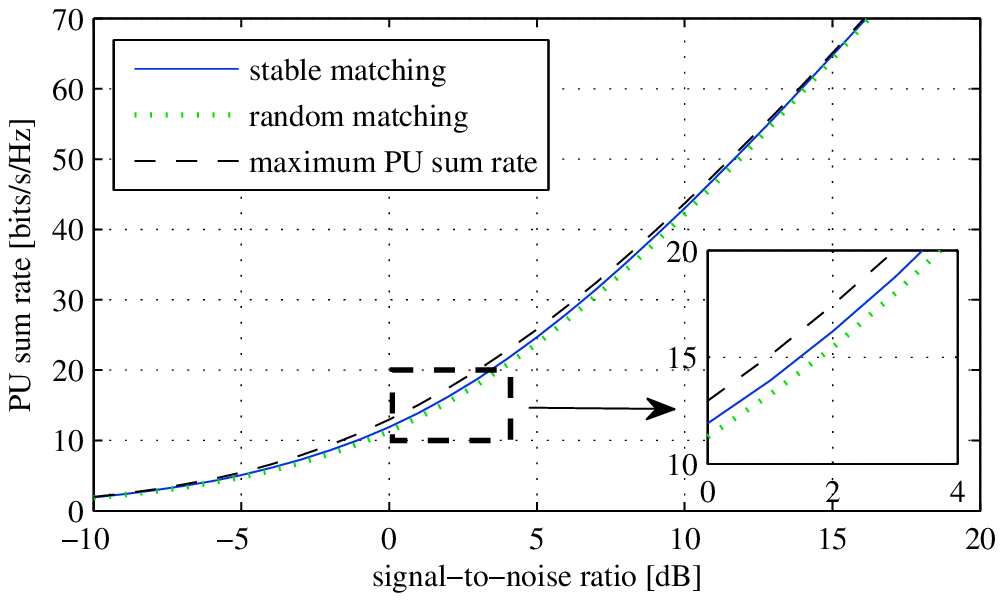}
	\caption{Average sum-rate of the PUs for increasing SNR and quotas $q_k=2$ for all $k$.}
	\label{fig:snr_pu_utility}
\end{figure}
\begin{figure}[t]
	\centering
	\includegraphics[width=\linewidth,clip]{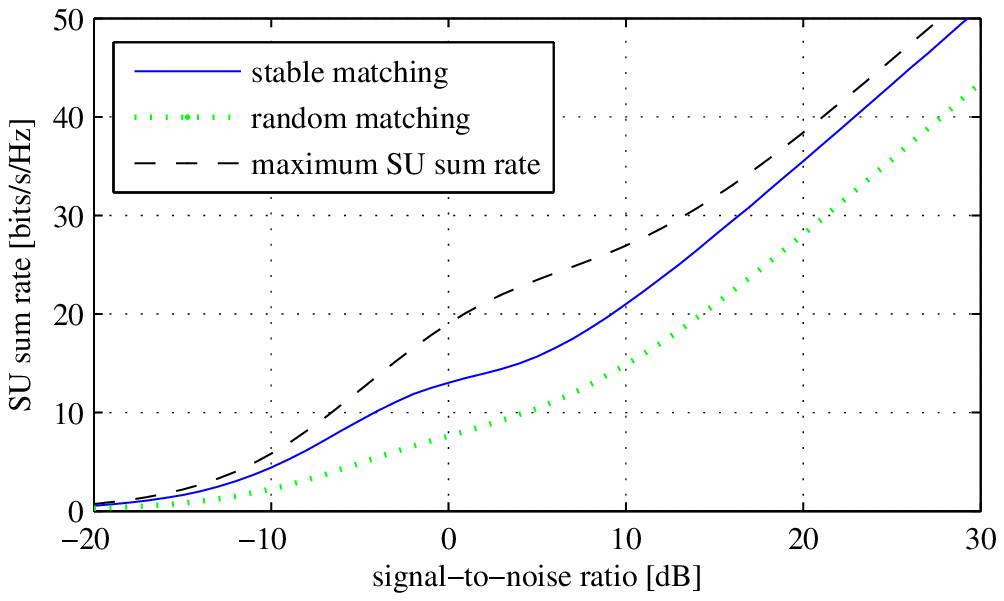}
	\caption{Average sum-rate of the SUs for increasing SNR and quotas $q_k=2$ for all $k$.}
	\label{fig:snr_su_utility}
\end{figure}

In \figurename~\ref{fig:region_snr0}(a), we plot the performance of stable matching following Algorithm \ref{alg:sm_su_optimal} for $q_k = 1$, for all $k \in \setK$. The outcome favours the SUs and is near the quoted boundary showing that the sum performance of the SUs from stable matching is near optimal. Note, that in this setting not all channels can be assigned to the SUs due to the quota restriction. In \figurename~\ref{fig:region_snr0}(b), we set $q_k = 2$, for all $k \in \setK$. In this setting, the outcomes of stable matching do not reach the boundary but are closer to it than the random matching scheme discussed further below. Note, that our proposed quick terminating algorithm leads to the best stable matching for the SUs but there may exist other stable matchings. Nevertheless, our stable matching outcome shows a fair trade-off in terms of giving an acceptable PU performance. In \figurename~\ref{fig:region_snr0}(b), we also plot the Walrasian equilibrium using Algorithm~\ref{alg:EnglishAuction} with price incrementing factor $\alpha = 0.005$ and weight $\lambda = 0.5$. By choosing $\lambda = 0.5$, we achieve in Walrasian equilibrium the maximum sum performance of the PUs and SUs in the quoted region. Note, that all points on the boundary of the quoted region can be obtained as Walrasian equilibria for different values of $\lambda$. In \figurename~\ref{fig:region_snr0}(c), the quotas are specified such that any SU can be assigned all primary channels. Following Theorem~\ref{thm:uniqueSM}, the stable matching is unique and is sum-rate optimal in the quoted region.

In comparison to stable matching for different quota values in \figurename~\ref{fig:region_snr0}(a-c), random matching chosen to satisfy the associated quota constraints does not show preference in performance to neither PUs nor SUs. In our random matching scheme, we first introduce $q_k$-many virtual SUs with a quota of one for each SU $k \in \setK$. Then, we apply random one-to-one matching achieving a number of $\min ( L, \sum_{k \in \setK}q_k )$ matching pairs.

%%%%%%%%%%%%%%%%%%%%%%%%%%%%%%%%%%%%%%%%%%%%%%%%%%%%%%%%%%%%%
% performance analysis of stable matching outcome vs. SNR
%%%%%%%%%%%%%%%%%%%%%%%%%%%%%%%%%%%%%%%%%%%%%%%%%%%%%%%%%%%%%
%\begin{figure}[t]
%	\centering
%	\includegraphics[width=\linewidth,clip]{figures/stabmatch_snr_proposals_pu_final.eps}
%	\caption{Average number of proposals from the coordinator, acting on behalf of the PUs, to the SUs for different quotas and increasing SNR occurring during Algorithm~\ref{alg:sm_pu_optimal}.}
%	\label{fig:snr_pu_utility2}
%\end{figure}
\begin{figure}[t]
	\centering
	\includegraphics[width=\linewidth,clip]{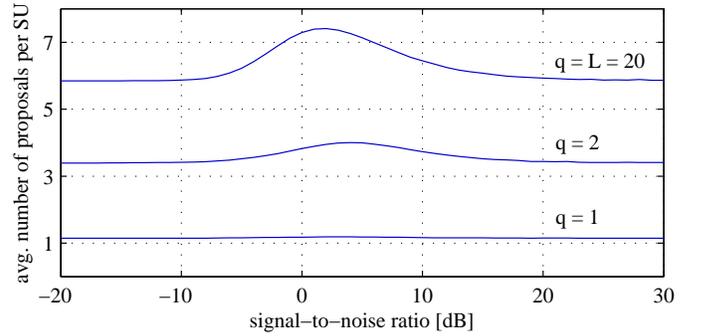}
	\caption{Average number of proposals from each SU to the coordinator for different quotas and increasing SNR occurring during Algorithm~\ref{alg:sm_su_optimal}.}
	\label{fig:snr_su_utility2}
\end{figure}
\subsection{Performance of Stable Matching}
In \figurename~\ref{fig:snr_pu_utility} and \figurename~\ref{fig:snr_su_utility}, the average sum rate of the PUs and SUs are plotted respectively for increasing SNR and for $q_k = 2$, for all $k \in \setK$. The number of SUs is $K=10$ and number of primary channels is $L=20$. All channels are independently and identically Rayleigh distributed and we generate again $10^3$ random instances for averaging in the simulations.
In \figurename~\ref{fig:snr_pu_utility}, it is shown that the performance loss of the PUs is very low in stable matching compared to the setting without the operation of SUs. Hence, the coexistence with the SUs does not lead to much performance degradations to the PUs.
In \figurename~\ref{fig:snr_su_utility}, the average sum rate of the SUs is shown to be always larger with Algorithm~\ref{alg:sm_su_optimal} than with the random matching scheme but does show a significant gap to the maximum possible sum rate of the SUs. Note, that the maximum SU sum rate is obtained at the point where the PU sum rate is lowest which is an unsatisfying operating point for a cognitive radio network, where primary user communication is prioritized. Hence, the SU sum rate reached by stable matching comes at an acceptable level. Since Algorithm~\ref{alg:sm_su_optimal} is SU-optimal, outcomes of other stable matching schemes would perform worse for the SUs. At high SNR, the sum rate performance of the SUs grows linearly. Note that in this SNR regime the detection probability of each SU approaches zero since the noise is much smaller compared to the primary signal power. Accordingly, the achievable rate of an SU is not limited by the interference from a PU.

\begin{figure}[t]
	\centering
	\includegraphics[width=\linewidth,clip]{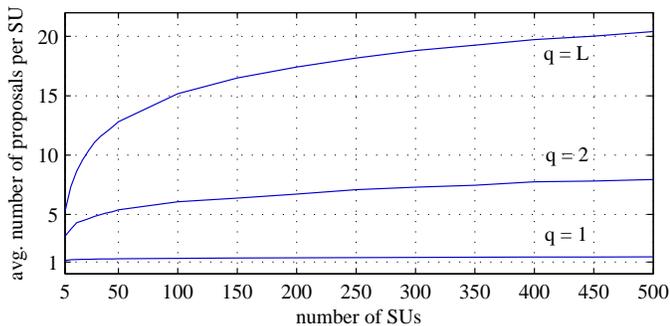}
	\caption{Average number of proposals by the SUs during Algorithm~\ref{alg:sm_su_optimal} at $0$ dB SNR. The number of PUs is double the number of SUs, and the quotas of the SUs are equal to $q$.}
	\label{fig:complexity_su}
\end{figure}

In \figurename~\ref{fig:snr_su_utility2}, the complexity of Algorithm~\ref{alg:sm_su_optimal} is revealed by counting the number of matching proposals per secondary user over SNR in the simulation scenario described above. The average number of SU proposals increases for larger quotas $q_k=q$, for all $k \in \setK$. For $q_k = 1$, for all $k \in \setK$, where the SUs require only one channel, the stable matching algorithm tends to match each SU with its first preference of the channels. Only few re-matchings occur. Hence, a very low average number of SU proposals (slightly above one) is needed for terminating the protocol. Re-matchings occur when a user, engaged to a channel, is released due to a proposal from another user which gives the channel a higher utility. For large $q_k$ requirements, frequent re-matchings lead to a much higher proposal number in the presented stable matching algorithm. Generally however, it is shown that only a few number of proposals are required to reach a stable matching. In \figurename~\ref{fig:complexity_su}, the average number of proposals during Algorithm~\ref{alg:sm_su_optimal} is plotted for increasing number of SUs. The number of PU channels is set to be double the number of SUs. It is shown that the average number of proposals during the stable matching algorithm increases with increasing quotas. 

%For the PU-optimal algorithm, the average number of proposals increases for smaller quotas.
%
%%%%%%%%%%%%%%%%%%%%%%%%%%%%%%%%%%%%%%%%%%%%%%%%%%%%%%%%%%%%%
% number of proposals per user vs. scaling factor
%%%%%%%%%%%%%%%%%%%%%%%%%%%%%%%%%%%%%%%%%%%%%%%%%%%%%%%%%%%%%
%\begin{figure}[t]
%	\centering
%	\includegraphics[width=\linewidth,clip]{figures/stabmatch_sunumber_proposals_su_final.eps}
%	\caption{Average number of proposals by the SUs during Algorithm~\ref{alg:sm_su_optimal} and by the coordinator, acting on behalf of the PUs during Algorithm~\ref{alg:sm_pu_optimal}, at $0$ dB SNR. The number of PUs is double the number of SUs, and the quotas of the SUs are equal to $q$.}
%	\label{fig:complexity_stabmatch}
%\end{figure}
%\begin{figure}[t]
%	\centering
%	\includegraphics[width=\linewidth,clip]{figures/stabmatch_sunumber_proposals_pu_final.eps}
%	\caption{Average number of proposals by the PUs during Algorithm~\ref{alg:sm_pu_optimal} at $0$ dB SNR. The number of PUs is double the number of SUs, and the quotas of the SUs are equal to $q$.}
%	\label{fig:complexity_pu}
%\end{figure}
%
%%%%%%%%%%%%%%%%%%%%%%%%%%%%%%%%%%%%%%%%%%%%%%%%%%%%%%%%%%%%%
% performance analysis for different alpha
%%%%%%%%%%%%%%%%%%%%%%%%%%%%%%%%%%%%%%%%%%%%%%%%%%%%%%%%%%%%%
\subsection{Performance of English Auction}
\begin{figure}[t]
	\centering
	\includegraphics[width=\linewidth,clip]{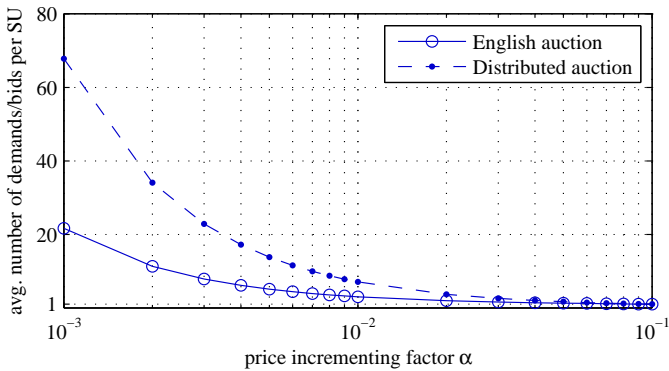}
	\caption{Comparison of average number of demands per user during Algorithm~\ref{alg:EnglishAuction} and the average number of bids from the distributed auction algorithm in \cite{Naparstek2014} for different values of $\alpha$ with $L = K = 10$ and $q_k = 1$ for all $k$. SNR is 0 dB.}
	\label{fig:demand_Walras}
\end{figure}

\begin{figure}[t]
	\centering
	\includegraphics[width=\linewidth,clip]{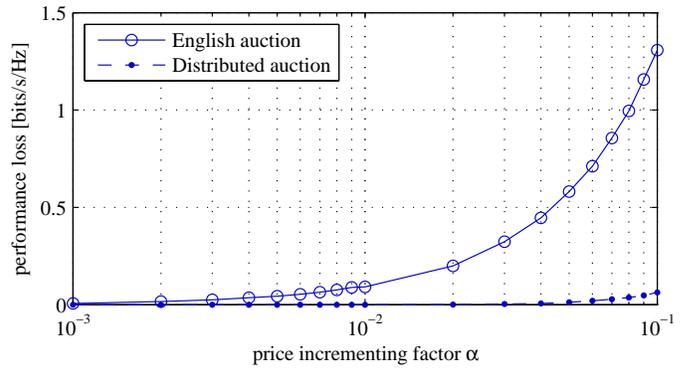}
	\caption{Comparison of performance loss in outcome of Algorithm~\ref{alg:EnglishAuction} and the distributed auction algorithm in \cite{Naparstek2014} for different values of $\alpha$ with $L = K = 10$ and $q_k = 1$ for all $k$. SNR is 0 dB.}
	\label{fig:perfloss_Walras}
\end{figure}

\begin{figure}[t]
	\centering
	\includegraphics[width=\linewidth,clip]{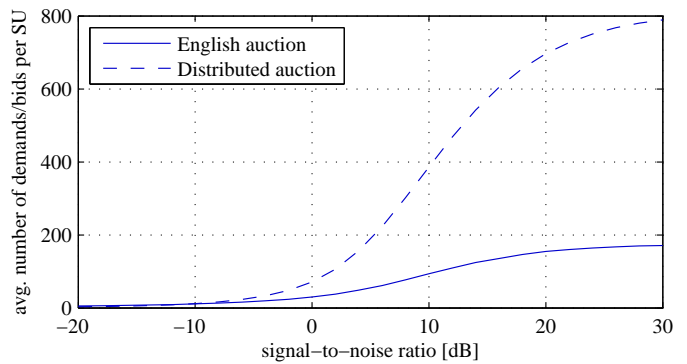}
	\caption{Comparison of average number of demands per user required for the auction algorithms for different SNR values with $\alpha = 0.01$ and $L = K = 10$ and $q_k = 1$ for all $k$.}
	\label{fig:demand_Walras_snr}
\end{figure}

\begin{figure}[t]
	\centering
	\includegraphics[width=\linewidth,clip]{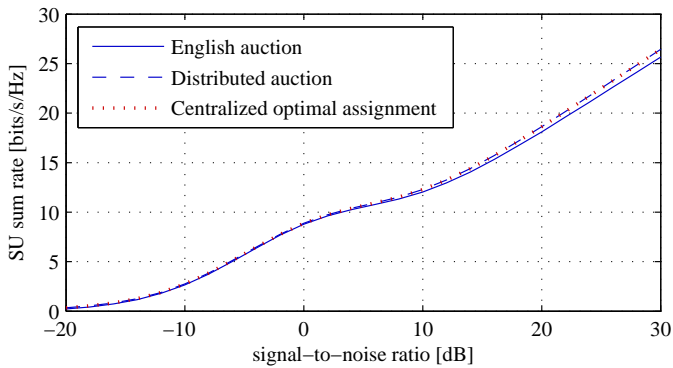}
	\caption{Comparison of average SU sum rate for different SNR values. The price incrementing factor $\alpha$ is $0.01$ and $L = K = 10$ and $q_k = 1$ for all $k$.}
	\label{fig:sumrate_Walras_snr}
\end{figure}

In \figurename~\ref{fig:demand_Walras} and \figurename~\ref{fig:perfloss_Walras}, we compare the average number of demands/bids and the performance loss from optimal channel assignment by Hungarian method approaches, respectively, with respect to the price incrementing factor $\alpha$ used in Algorithm~\ref{alg:EnglishAuction} and also required in the distributed auction mechanism in \cite[Section IV]{Naparstek2014}. Note that in \cite[Section V]{Naparstek2014}, an implementation of the distributed auction mechanism is provided using opportunistic CSMA which needs no exchange of information between the users. In order to compare the performance to the algorithm in \cite{Naparstek2014} which assigns a single channel per user, we set $q_k = 1$, for all $k \in \setK$ and choose $L = K = 10$. We average the performance at SNR $= 0$ dB over $10^3$ channel realizations. In \figurename~\ref{fig:demand_Walras}, it can be seen that the auction algorithm in \cite{Naparstek2014} requires on average larger numbers of bids per SU compared to the number of demands per user in the English auction. However, according to \figurename~\ref{fig:perfloss_Walras}, the English auction is shown to be more sensitive over the choice of $\alpha$. A larger $\alpha$ increases the convergence rate of the English auction, but leads to high performance loss. The low sensitivity of the distributed auction in \cite{Naparstek2014} to the price incrementing factor is due to the fact that this auction method makes use of the prices of both the best and second best object a user would demand to determine his bid. This cannot be exploited in the English auction since the demand of a user is a \emph{set} of channels.

In \figurename~\ref{fig:demand_Walras_snr} and \figurename~\ref{fig:sumrate_Walras_snr}, we plot the average number of demands and the SU sum rate for different SNR values. The chosen price increment factor for the auction algorithms is $\alpha = 0.01$. It can be seen that the distributed auction requires more average number of bids for larger SNR values than the English auction requires number of demands. Both algorithms achieve very close performance to the optimum as is shown in \figurename~\ref{fig:sumrate_Walras_snr}. The optimum is reached centralized by means of the Hungarian method.
Note that the performance loss of both algorithms is due to the non-infinitesimal chosen value of the price incrementing factor.%
%
%
%I would stress the relation of price incrementing factor to convergence time of the algorithm and say that complexity is an issue in finding the Walrasian eq.
%
%\begin{figure}[!h]
%	\centering
%	%\includegraphics[width=9cm,clip]{figures/perfloss_Walras.eps}
%	\caption{Plot of average number of demands per user during Algorithm~\ref{alg:EnglishAuction} for different SNR values with $L = K = 10$ and $q_k = 1$ for all $k$.}
%	\label{fig:demand_Walras_SNR}
%\end{figure}
%\begin{figure}[!h]
%	\centering
%	%\includegraphics[width=9cm,clip]{figures/perfloss_Walras.eps}
%	\caption{Plot of performance loss in the outcome of Algorithm~\ref{alg:EnglishAuction} for different SNR values with $L = K = 10$ and $q_k = 1$ for all $k$.}
%	\label{fig:perfloss_Walras_SNR}
%\end{figure}
%
%%%%%%%%%%%%%%%%%%%%%%%%%%%%%%%%%%%%%%%
%%%%%%%%%%%%%%%%%%%%%%%%%%%%%%%%%%%%%%%
%\begin{figure*}%[!ht]
%
%    \subfloat[quotas $q_k=1 \, \forall \, k$ \label{fig:region_snr0_q1}]{%
%      \includegraphics[height=3.9cm]{figures/region/region_snr10_quota1.eps}
%    }
%    \hfill
%    \subfloat[quotas $q_k=2 \, \forall \, k$ \label{fig:region_snr0_q2}]{%
%      \includegraphics[height=3.9cm]{figures/region/region_snr10_quota2.eps}
%    }
%    \hfill
%    \subfloat[quotas $q_k=L \, \forall \, k$ \label{fig:region_snr0_q20}]{%
%      \includegraphics[height=3.9cm]{figures/region/region_snr10_quota20.eps}
%    }
%
%    \caption{Matching Performance: Utility Region at 10 dB SNR}
%	\label{fig:region_snr10}
%
%\end{figure*}

\section{Conclusion}\label{sec:conclusions}

We considered the problem of assigning primary channels to SUs for communication in a cognitive radio. For this problem, we proposed two solution concepts, stable matching and Walrasian equilibrium, and provided coordination and cooperation algorithms to reach them in a distributed manner. Both concepts lead to different performances for the secondary and primary users. We relate both solutions utilizing the achievable sum performance regions. While stable matching relies on stability of the assigned secondary and primary users, the Walrasian equilibrium maximizes the weighted sum utilities. In contrast to the stable matching framework, prices are required in the competitive market model in order to define the Walrasian equilibrium.

The complexity of SU-optimal stable matching algorithms in terms of the average number of proposals is shown to be very low by extensive simulation. The complexity of the English auction to reach a Walrasian equilibrium depends, however, on the choice of the price incrementing parameter. Future works may devise a mechanism which adapts the price incrementing parameter intelligently.

\appendix
\subsection{Proof of Theorem \ref{thm:uniqueSM}}\label{proof:uniqueSM}
The proof is based on the result in \cite{Roth1986} (also given in \cite[Theorem 5.13]{Roth1990}) which we restate here in relation to our cognitive radio setting.
\begin{theorem}[{\cite{Roth1986}}]\label{thm:roth}
Let $k$ be a secondary user with quota $q_k$, and let $M$ be a stable matching such that $\abs{M(k) \cap \setL} < q_k$. Then for any stable matching $M'$, $M'(k) = M(k)$.
\end{theorem}
Thus, if a stable matching does not strictly satisfy the quota of a secondary user with equality, then this user is assigned the same set of primary channels in any other stable matching. Since $q_k = L$ for all $k\in \setK$, and the existence of at least one stable matching is guaranteed, then we can do the following two-case study: 1) If $\abs{M(k)} < q_k$ for all $k \in \setK$, then the stable matching is unique Theorem \ref{thm:roth} holds for all users. 2) If $\abs{M(k)} = q_k$ for some user $k \in \setK$ then $\abs{M(j)} = 0$ for all $j \neq k$. Following Theorem \ref{thm:roth} all users $j \neq k$ are unmatched in all other stable matchings. What is left is to consider stable matchings in which the user $k$ is assigned strictly less channels than his quota. However, since a matching $M'$, with $\abs{M'(k)} < q_k$ is not individually rational for user $k$ according to Definition \ref{def:individually_rational}, then $M'$ is not a stable matching. Hence, $M$ is unique.

Considering Algorithm \ref{alg:sm_su_optimal}, each SU $k$ proposes to every channel since his quota $q_k \geq L$. The coordinator then accepts the proposal of an SU $k$ in a channel $l$ if he gives the highest performance for PU $l$. Accordingly, the unique stable matching is sum performance optimal for the PUs within the performance region $\widetilde{\mathcal{R}}^\pu$ defined in \eqref{eq:rate_region_PU_tilde}.%
\subsection{Proof of Theorem \ref{thm:existence}}\label{proof:existence}
We have to show that the utility function in \eqref{eq:utility_quota} satisfies the monotonicity property and the gross substitutes condition. The monotonicity property is obviously satisfied because if additional channels are provided to an SU $k$, the utility $U_k$ in \eqref{eq:utility_quota} does not decrease. In order to prove that the utility function in \eqref{eq:utility_quota} satisfies the gross substitutes condition, we must first prove that
\begin{equation}\label{eq:add_sep}
\phi_k(\setB) = \lambda u^\susum_k(\setB) + (1 - \lambda) \sum\nolimits_{l \in \setB} u^\pu_l(k),
\end{equation}
satisfies the gross substitute property. Afterwards, the operation in \eqref{eq:utility_quota} on $\phi_k(\setB)$ is called the $q_k$-\emph{satiation} of $\phi_k$ and preserves the gross substitutes property according to \cite[Section 2]{Gul1999}. The function $\phi_k(\setB)$ satisfies the gross substitutes property because it is \emph{additively separable} \cite[Section 2]{Gul1999}, i.e., $\phi_k(\setB)$ in \eqref{eq:add_sep} can be expressed as $\phi_k(\setB) = \sum\nolimits_{l \in \setB} \phi_k(\{l\})$.

\subsection{Proof of Lemma \ref{lem: quotas_not_violated}}\label{proof: quotas_not_violated}
The proof is by contradiction. Given $p_l>0$ for all $l \in \setL$, assume for some SU $k$ a demand set $\setA \in \setD_k(\mat{p})$ satisfies $\abs{\setA} > q_k$. Then, according to his utility function in \eqref{eq:utility_quota}, a set of channels $\setR \subset \setA$ with $\abs{\setR} \geq \abs{\setA} - q_k$ give no additional performance to SU $k$ since the maximization in \eqref{eq:utility_quota} is over at most $q_k$ channels and we can write $U_k(\setA) = U_k(\setA \setminus \setR)$. The net utility of SU $k$ in \eqref{eq:net_utility} satisfies
\begin{subequations}
\begin{align}
v_k(\setA, \mat{p}) & = U_k(\setA) - \sum\limits_{l \in \setA} p_l \\
&= U_k(\setA \setminus \setR) - \sum\limits_{l \in \setA\setminus \setR} p_l - \sum\nolimits_{l' \in \setR} p_{l'}\\
&= v_k(\setA \setminus \setR, \mat{p}) - \sum\limits_{l' \in \setR} p_{l'}.
\end{align}
\end{subequations}
\noindent Hence, $v_k(\setA, \mat{p}) < v_k(\setA \setminus \setR, \mat{p})$ which contradicts that $\setA$ is a demand for SU $k$.

\subsection{Proof of Theorem \ref{thm:greedy_demand}}\label{proof:greedy_demand}

The consumer utility function in \eqref{eq:utility_quota} is proven in Theorem \ref{thm:existence} to satisfy the gross substitutes condition. We will use an equivalent property to the gross substitutes condition called the single improvement property \cite{Gul1999} in the proof. 
\begin{definition}\label{def:single_improvement}
The utility function $U_k(\setA)$ satisfies the \emph{single improvement property} if for any price vector $\mat{p}$ and set of channels $\setA \neq \setD_k(\mat{p})$, i.e., $\setA$ is not a demand set for SU $k$, there exists another set $\setB$ such that $v_k(\setA,\mat{p}) < v_k(\setB,\mat{p})$ with either $\abs{\setA\setminus \setB} \leq 1$, or $\abs{\setB\setminus \setA} \leq 1$.
\end{definition}

Let $\setA_k$ be the output of Algorithm~\ref{alg:demand} and assume $\setA_k$ is not a demand set, i.e., $\setA_k \notin \setD_k(\mat{p})$. Then in order to strictly improve $v_k(\setA_k,\mat{p})$ we can, according to the single improvement property in Definition \ref{def:single_improvement}, do either of three possibilities:
\begin{description}
\item[(i)] adding an element to $\setA_k$, i.e., find $a \in \setL \setminus \setA_k$ s.t. $v_k(\setA_k \cup \{a\},\mat{p}) > v_k(\setA_k,\mat{p})$
\item[(ii)] removing one element from $\setA_k$, i.e., find $b \in \setL \setminus \setA_k$ s.t. $v_k(\setA_k \setminus \{b\},\mat{p}) > v_k(\setA_k,\mat{p})$
\item[(iii)] or do both (i) and (ii).%, i.e., find $a \in \setL \setminus \setA_k$ and $b \in \setL \setminus \setA_k$ such that $v_k(\setA_k \setminus \{b\} \cup \{a\},\mat{p}) > v_k(\setA_k,\mat{p})$.
\end{description}

%The utility function $U_k$ satisfies the single improvement property in Definition \ref{def:single_improvement}. 

Case (i): If $\abs{\setA_k} = q_k$, then adding an element $a \in \setL \setminus \setA_k$ requires removing another element $b\in \setA_k$ following Lemma \ref{lem: quotas_not_violated}. Since the elements in $\setA_k$ are the first $q_k$ best channels, then removing one element to add the element $a$ does not lead to strict performance improvement having that the net utilities in the resources are distinct according to Assumption~\ref{ass:unique_demand}. If $\abs{\setA_k} < q_k$, then the net utility with channel $a \in \setL \setminus \setA_k$ is nonpositive, i.e., $v_k(a,p_a) \leq 0$ because otherwise it would be a member of $\setA_k$ from Algorithm~\ref{alg:demand}. Hence, no strict performance improvement can be made by adding an element to $\setA_k$.

Case (ii): Since $v_k(b,p_b) > 0$ for all $b \in \setA_k$, then removing one element from $\setA_k$ leads to strict performance degradation.

Since both (i) and (ii) lead to no strict performance improvement, then (iii) cannot be satisfied. Accordingly, $\setA_k$ is a demand set, i.e., belongs to the set $\setD_k(\mat{p})$. From (ii), the $\setA_k$ is the smallest demand set in $\setD_k(\mat{p})$ and is contained in other demand sets in $\setD_k(\mat{p})$. Specifically, another demand set can be constructed from $\setA_k$ when $\abs{\setA_k} < q_k$ by adding a resource $l \in \setL \setminus \setA_k$ to $\setA_k$ which satisfies $v_k(\{l\},\mat{p}) = 0$.

\subsection{Proof of Theorem \ref{thm:excess_demand}}\label{proof:excess_demand}
Following Theorem \ref{thm:greedy_demand}, the demand is the smallest element in $\setD_k(\mat{p})$ in \eqref{eq:demand}. Then, for an SU $k$, the intersection of his demand set $\setA_k$ from Algorithm~\ref{alg:demand} with the aggregate excess demand $\setZ$ from Algorithm~\ref{alg:excess_demand} is smallest compared to other demand sets in $\setD_k(\mat{p})$.

First, we prove that $\setZ$ belongs to the set $\mathcal{O}(\mat{p})$ in \eqref{eq:max_dem}. That is, we must prove:
\begin{description}
\item[(i)] $K_{\setK} (\setZ,\mat{p}) - \abs{\setZ} = K_{\setK} (\setZ \cup \{a\},\mat{p}) - \abs{\setZ \cup \{a\}}$, $a \in \setL \setminus \setZ$
\item[(ii)] $K_{\setK} (\setZ,\mat{p}) - \abs{\setZ} > K_{\setK} (\setZ \setminus \{b\},\mat{p}) - \abs{\setZ \setminus \{b\}}$, $b \in \setZ$.
\end{description}
In order to prove (i), let the element $a \in \setL \setminus \setZ$ be demanded by SU $j$ only, i.e., $ a \in \setA_j$. Note, that if channel $a$ is demanded by more than two SUs, then it would belong to the set $\setZ$ according to Algorithm ~\ref{alg:excess_demand}. Then we can write
\begin{subequations}
\begin{align}
& K_{\setK} (\setZ \cup \{a\},\mat{p}) - \abs{\setZ \cup \{a\}} \\
& = \sum\nolimits_{k=1}^K \abs{\setA_k \cap \setZ \cup \{a\}} - \abs{\setZ \cup \{a\}} \\
& = \sum\nolimits_{k=1, k\neq j}^K \abs{\setA_k \cap \setZ} + \abs{\setA_j \cap \setZ \cup \{a\}} - \abs{\setZ} - 1\\
& = \sum\nolimits_{k=1, k\neq j}^K \abs{\setA_k \cap \setZ} + \abs{\setA_j \cap \setZ} + 1 - \abs{\setZ} - 1 \\
& = \sum\nolimits_{k=1}^K \abs{\setA_k \cap \setZ} - \abs{\setZ}.
\end{align}
\end{subequations}
In order to prove (ii), let $b \in \setZ$ be demanded by SUs $j$ and $\ell$, i.e., $b \in \setA_j$ and $b \in \setA_\ell$. Then,
\begin{subequations}
\begin{align}
& K_{\setK} (\setZ,\mat{p}) - \abs{\setZ} = \sum\limits_{k=1}^K \abs{\setA_k \cap \setZ} - \abs{\setZ} \\
& = \sum\limits_{\substack{k = 1 \\ k \notin \{j,\ell\}}}^K \abs{\setA_k \cap \setZ} + \abs{\setA_j \cap \setZ} + \abs{\setA_\ell \cap \setZ} - \abs{\setZ}\\ \nonumber
& = \sum\nolimits_{k=1,k \notin\{j,\ell\}}^K \abs{\setA_k \cap \setZ \setminus \{b\} }+ \abs{\setA_j \cap \setZ\setminus \{b\}}  \\ & \quad + \abs{\setA_\ell \cap \setZ \setminus \{b\}} + 2 - \abs{\setZ \setminus \{b\}} - 1\\
& = \sum\nolimits_{k=1}^K \abs{\setA_k \cap \setZ \setminus \{b\}} - \abs{\setZ \setminus \{b\}} + 1\\ & > K_{\setK} (\setZ \setminus \{b\},\mat{p}) - \abs{\setZ \setminus \{b\}}.
\end{align}
\end{subequations}
Thus, $\setZ \in \mathcal{O}(\mat{p})$ and from (ii), $\setZ$ is the smallest element in $\mathcal{O}(\mat{p})$.

\bibliographystyle{IEEEtran}%
\bibliography{collection}
\end{document}